\documentclass[acmsmall]{acmart}
\usepackage{multirow}
\usepackage{subfigure}
\usepackage{amsmath}
\usepackage{algorithm}
\usepackage{algorithmic}

\AtBeginDocument{%
  \providecommand\BibTeX{{%
    \normalfont B\kern-0.5em{\scshape i\kern-0.25em b}\kern-0.8em\TeX}}}

\setcopyright{acmcopyright}
\acmYear{2022}
\acmDOI{10.1145/3569423}

\acmJournal{TIST}
\acmVolume{1}
\acmNumber{1}
\acmArticle{1}
\acmMonth{10}
\acmPrice{15.00}

\begin{document}

\title[Learning to Rank Explanations]{On the Relationship between Explanation and Recommendation: Learning to Rank Explanations for Improved Performance}


\author{Lei Li}
\affiliation{%
	\institution{Hong Kong Baptist University}
	\streetaddress{34 Renfrew Road}
	\city{Hong Kong}
	\country{China}
}
\email{csleili@comp.hkbu.edu.hk}

\author{Yongfeng Zhang}
\affiliation{%
	\institution{Rutgers University}
	\streetaddress{110 Frelinghuysen Road}
	\city{New Brunswick}
	\state{New Jersey}
	\country{USA}
	\postcode{08854-8019}
}
\email{yongfeng.zhang@rutgers.edu}

\author{Li Chen}
\affiliation{%
	\institution{Hong Kong Baptist University}
	\streetaddress{34 Renfrew Road}
	\city{Hong Kong}
	\country{China}
}
\email{lichen@comp.hkbu.edu.hk}

\renewcommand{\shortauthors}{Li, Zhang, and Chen}

\begin{abstract}
Explaining to users why some items are recommended is critical, as it can help users to make better decisions, increase their satisfaction, and gain their trust in recommender systems (RS). However, existing explainable RS usually consider explanation as a side output of the recommendation model, which has two problems: (1) it is difficult to evaluate the produced explanations because they are usually model-dependent, and (2) as a result, how the explanations impact the recommendation performance is less investigated.

In this paper, explaining recommendations is formulated as a ranking task, and learned from data, similar to item ranking for recommendation. This makes it possible for standard evaluation of explanations via ranking metrics (e.g., NDCG). Furthermore, this paper extends traditional item ranking to an item-explanation joint-ranking formalization to study if purposely selecting explanations could reach certain learning goals, e.g., improving recommendation performance. A great challenge, however, is that the sparsity issue in the user-item-explanation data would be inevitably severer than that in traditional user-item interaction data, since not every user-item pair can be associated with all explanations. To mitigate this issue, this paper proposes to perform two sets of matrix factorization by considering the ternary relationship as two groups of binary relationships. Experiments on three large datasets verify the solution's effectiveness on both explanation ranking and item recommendation.
\end{abstract}



\maketitle

\section{Introduction}

Recommendation algorithms, such as collaborative filtering \cite{CSCW94-UCF, WWW01-ICF} and matrix factorization \cite{NIPS08-PMF, Computer09-MF}, have been widely deployed in online platforms, such as e-commerce and social networks, to help users find their interested items.
Meanwhile, there is a growing interest in explainable recommendation \cite{SIGIR14-EFM, ICDM18-RL, AAAI19-DER, WWW18-NARRE, AAAI19-DEAML, IJCAI19-CAML, SIGIR17-NRT, CIKM20-NETE, SIGIR19-VECF, CIKM15-TriRank, FTIR20-Survey}, which aims at producing user-comprehensible explanations, as they can help users make informed decisions and gain users' trust in the system \cite{Handbook15-Explanation, FTIR20-Survey}.
However, in current explainable recommendation approaches, explanation is often a side output of the model, which would incur two problems:
first, the standard evaluation of explainable recommendation could be difficult, because the explanations vary from model to model (i.e., model-dependent);
second, these approaches rarely study the potential impacts of explanations, mainly because of the first problem.

Evaluation of explanations in existing works can be generally classified into four categories, including case study, user study, online evaluation and offline evaluation \cite{FTIR20-Survey}.
In most works, case study is adopted to show how the example explanations are correlated with recommendations.
These examples may look intuitive, but they are less representative to reflect the overall quality of the explanations.
Results of user study \cite{balog2020measuring, WWW21-ELIXIR} are more plausible, but it can be expensive and is usually evaluated in simulated environments which may not reflect real users' actual perception.
Though this is not a problem in online evaluation, it is difficult to implement as it relies on the collaboration with industrial firms, which may explain why only few works \cite{SIGIR14-EFM, WWW20-Office, RecSys18-BART} conducted online evaluation.
Consequently, one may wonder whether it is possible to evaluate the explainability using offline metrics.
However, as far as we know, there is no standard metrics that are well recognized by the community.
Though BLEU \cite{ACL02-BLEU} and ROUGE \cite{TS04-ROUGE} have been widely adopted to evaluate text quality for natural language generation, text quality is not equal to explainability \cite{CIKM20-NETE, EARS19-HSS}.

\begin{figure}
	\centering
	\includegraphics[scale=0.5]{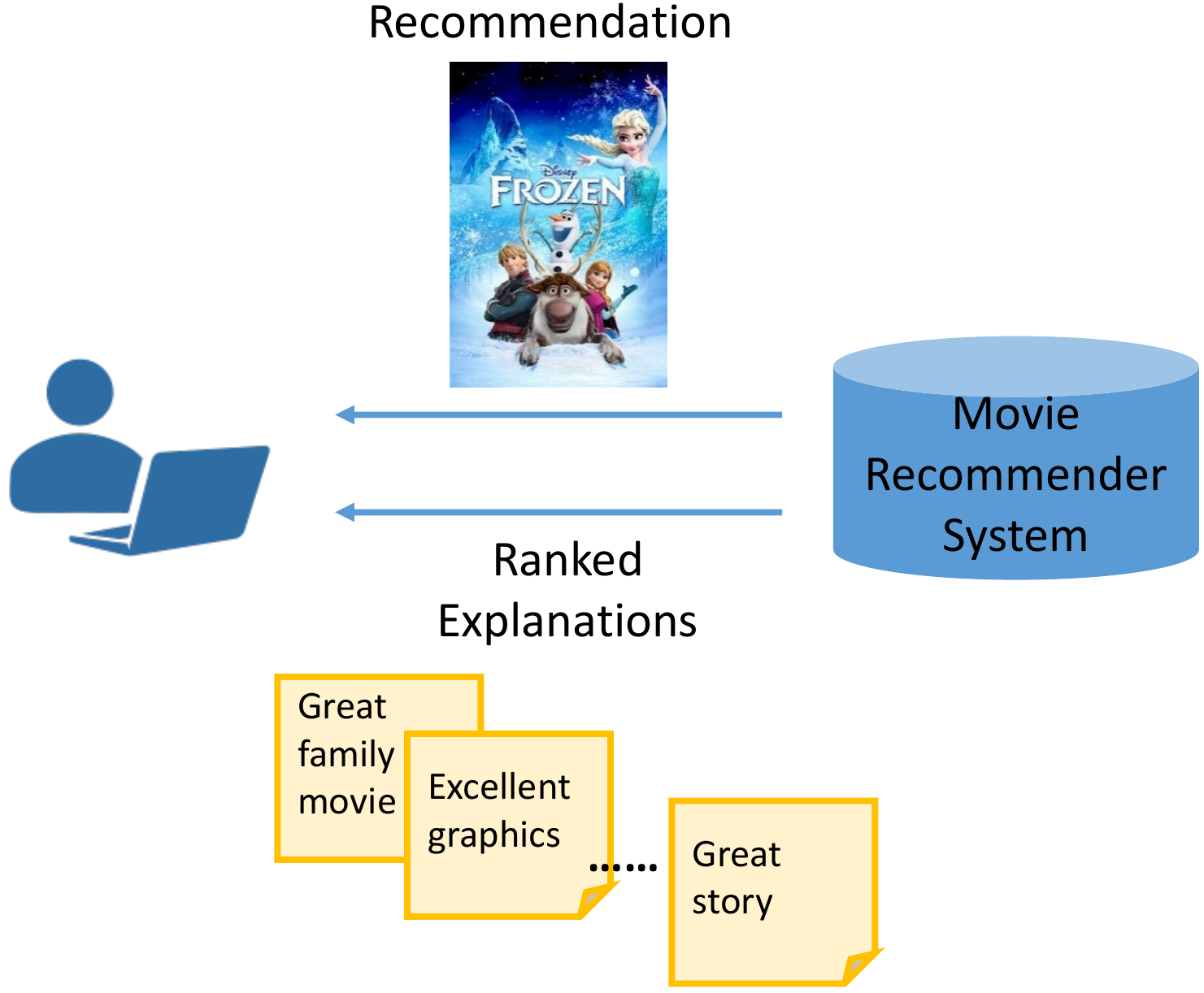}
	\caption{A toy example of explanation ranking for a movie recommender system.}
	\label{fig:example}
\end{figure}

With the attempt to achieve a standard offline evaluation of recommendation explanations, we formulate the explanation problem as a ranking task \cite{Book11-L2R}.
The basic idea is to train a model that can select appropriate explanations from a shared explanation pool for a recommendation.
For example, when a movie recommender system suggests the movie ``Frozen'' to a user, it may also provide a few explanations, such as ``great family movie'' and ``excellent graphics'', as shown in Fig. \ref{fig:example}.
Notice that, these explanations are available all the time, but their ranking orders differ for different movie recommendations, and only those ranked top are presented to the user.
In this case, the explanations are also learned from data, similar to recommendations.
Moreover, this general formulation can be adapted to various explanation styles, such as sentences, images, and even new styles yet to be invented, as long as the user-item-explanation interactions are available.
As an instantiation, we adopt three public datasets with textual explanations \cite{SIGIR21-EXTRA} for experimentation.

With the evaluation and data, we can investigate the potential impacts of explanations, such as higher chance of item click, conversion or fairness \cite{singh2018fairness}, which are less explored but are particularly important in commercial systems.
Without an appropriate approach to explanation evaluation, explanations have usually been modeled as an auxiliary function of the recommendation task in most explainable models \cite{SIGIR14-EFM, WWW18-NARRE, RecSys17-D-Attn, WWW18-TARMF, AAAI19-DER}.
Recent works which jointly model recommendation and text generation \cite{IJCAI19-CAML} or feature prediction \cite{AAAI19-DEAML, SIGIR18-MTER}, find that the two tasks could influence each other.
In particular, \cite{SIGIR16-LRPPM} shows that fine-tuning the parallel task of feature ranking can boost the recommendation performance.
Moreover, a user study shows that users' feedback on explanation items could help to improve recommendation accuracy \cite{WWW21-ELIXIR}.
Based on these findings, we design an item-explanation joint-ranking framework to study if showing some particular explanations could lead to increased item acceptance rate (i.e., improving the recommendation performance).
Furthermore, we are motivated to identify how the recommendation task and the explanation task would interact with each other, whether there is a trade-off between them, and how to achieve the most ideal solution for both.

However, the above investigation cannot proceed without addressing the inherent data sparsity issue in the user-item-explanation interactions.
In traditional pair-wise data, each user may be associated with several items, but in the user-item-explanation triplets data, each user-item pair may be associated with only one explanation.
In consequence, the data sparsity problem is severer for explanation ranking.
Therefore, how to design an effective model for such one-shot learning scenario becomes a great challenge.
Our solution is to separate user-item-explanation triplets into user-explanation and item-explanation pairs, which significantly alleviates the data sparsity problem.
Based on this idea, we design two types of model.
First, a general model that only makes use of IDs, aims to accommodate a variety of explanation styles, such as sentences and images.
Second, a domain-specific model based on BERT \cite{NAACL19-BERT} further leverages the semantic features of the explanations to enhance the ranking performance.

In summary, our key contributions are as follows:
\begin{itemize}
	\item To the best of our knowledge, our work is the first attempt to achieve standard evaluation of explainability for explainable recommendation via well-recognized metrics, such as NDCG, precision and recall.
	We realize this by formulating the explanation problem as a ranking-oriented task.
	\item With the evaluation, we further propose an item-explanation joint-ranking framework that can reach our designed goal, i.e., improving the performance of both recommendation and explanation, as evidenced by our experimental results.
	\item To that end, we address the data sparsity issue in the explanation ranking task by designing an effective solution, being applied to two types of model (with and without semantic features of the explanations)\footnote{Codes available at https://github.com/lileipisces/BPER}.
	Extensive experiments show their effectiveness against strong baselines.
\end{itemize}

In the following, we first summarize related work in Section \ref{sec:related}, and then formulate the problems in Section \ref{sec:problem}.
Our proposed models and the joint-ranking framework are presented in Section \ref{sec:model}.
Section \ref{sec:setup} introduces the experimental setup, and the discussion of results is provided in Section \ref{sec:results}.
We conclude this work with outlooks in Section \ref{sec:conclude}.

\section{Related Work} \label{sec:related}

Recent years have witnessed a growing interest in explainable recommendation \cite{SIGIR14-EFM, AAAI19-DEAML, IJCAI19-CAML, SIGIR17-NRT, SIGIR19-VECF, CIKM20-NETE, RecSys17-D-Attn, WWW18-TARMF, WWW18-NARRE, RecSys17-TransNets, AAAI19-DER, ICDM18-RL, JIIS21-CAESAR}.
In these works, there is a variety of explanation styles to recommendations, including visual highlights \cite{SIGIR19-VECF}, textual highlights \cite{RecSys17-D-Attn, WWW18-TARMF}, item neighbors \cite{WWW21-ELIXIR}, knowledge graph paths \cite{xian2019reinforcement, chen2021temporal, huang2021path}, word cloud \cite{SIGIR14-EFM}, item features \cite{CIKM15-TriRank}, pre-defined templates \cite{SIGIR14-EFM, AAAI19-DEAML, JIIS21-CAESAR}, automatically generated text \cite{IJCAI19-CAML, SIGIR17-NRT, CIKM20-NETE, ACL21-PETER, yang2021explanation, yang2022comparative}, retrieved text \cite{WWW18-NARRE, RecSys17-TransNets, AAAI19-DER, ICDM18-RL, wang2022graph}, etc.
The last type of style is related to this paper, but explanations in these works are merely side outputs of their recommendation models.
As a result, none of these works measured the explanation quality based on benchmark metrics.
In comparison, we formulate the explanation task as a learning to rank \cite{Book11-L2R} problem, which enables standard offline evaluation via ranking-oriented metrics.

On the one hand, the application of learning to rank can also be found in other domains.
For instance,
\cite{ACL15-KG, IJCAI17-Baidu} attempt to explain entity relationships in Knowledge Graphs.
The major difference from our work is that they heavily rely on the semantic features of explanations, either constructed manually \cite{ACL15-KG} or extracted automatically \cite{IJCAI17-Baidu}, while one of our models works well when leveraging only the relation of explanations to users and items, without considering such features.

On the other hand, the appropriateness of current evaluation for explanations is still under debate.
There are some works \cite{IJCAI19-CAML, SIGIR17-NRT} that regard text similarity metrics (i.e., BLEU \cite{ACL02-BLEU} in machine translation and ROUGE \cite{TS04-ROUGE} in text summarization) as explainability, when generating textual reviews/tips for recommendations.
However, text similarity does not equal to explainability \cite{CIKM20-NETE, EARS19-HSS}.
For example, when the ground-truth is ``\textit{sushi is good}'', two generated explanations ``\textit{ramen is good}'' and ``\textit{sushi is delicious}'' gain the same score on the two metrics.
However, from the perspective of explainability, the latter is obviously more related to the ground-truth, as they both refer to the same feature ``\textit{sushi}'', but the metrics fail to reflect this issue.
As a response, in this paper we propose a new evaluation approach based on ranking.

Our proposed models are experimented on textual datasets, but it can be applied to a broad spectrum of other explanation styles, e.g., images, as discussed earlier.
Concretely, on each dataset there is a pool of candidate explanations to be selected for each user-item pair.
A recent online experiment \cite{WWW20-Office} conducted on Microsoft Office 365\footnote{https://www.office.com} shows that this type of globally shared explanations is indeed helpful to users.
The main focus of this work is to study how users perceive explanations, which is different from ours that aims to design effective models to rank explanations.
Despite of that, their research findings motivate us to provide better explanations that could lead to improved recommendations.

In more details, we model the user-item-explanation relations for both item and explanation ranking.
There is a previous work \cite{CIKM15-TriRank} that similarly considers user-item-aspect relations as a tripartite graph, where aspects are extracted from user reviews.
Another branch of related work is tag recommendation for folksonomy \cite{WSDM10-PITF, PKDD07-FolkRank}, where tags are ranked for each given user-item pair.
In terms of problem setting, our work is different from the preceding two, because they solely rank either items/aspects \cite{CIKM15-TriRank} or tags \cite{WSDM10-PITF, PKDD07-FolkRank}, while besides that we also rank item-explanation pairs as a whole in our joint-ranking framework.
Another difference is that we study how semantic features of explanations could help enhance the performance of explanation ranking, while none of them did so.

\begin{table}
	\centering
	\caption{Key notations and concepts.}
	\begin{tabular}{l|l}
		\hline
		\textbf{Symbol} & \textbf{Description} \\
		\hline
		$\mathcal{T}$ & training set \\
		$\mathcal{U}$ & set of users \\
		$\mathcal{I}$ & set of items \\
		$\mathcal{I}_u$ & set of items that user $u$ preferred \\
		$\mathcal{E}$ & set of explanations \\
		$\mathcal{E}_u$ & set of user $u$'s explanations \\
		$\mathcal{E}_i$ & set of item $i$'s explanations \\
		$\mathcal{E}_{u, i}$ & set of explanations that user $u$ preferred w.r.t. item $i$ \\
		\hline
		$\mathbf{P}$ & latent factor matrix for users  \\
		$\mathbf{Q}$ & latent factor matrix for items  \\
		$\mathbf{O}$ & latent factor matrix for explanations  \\
		$\mathbf{p}_u$ & latent factors of user $u$  \\
		$\mathbf{q}_i$ & latent factors of item $i$  \\
		$\mathbf{o}_e$ & latent factors of explanation $e$  \\
		$b_i$ & bias term of item $i$ \\
		$b_e$ & bias term of explanation $e$\\
		\hline
		$d$ & dimension of latent factors \\
		$\alpha$, $\lambda$ & regularization coefficient \\
		$\gamma$ & learning rate \\
		$T$ & iteration number \\
		$M$ & number of recommendations for each user \\
		$N$ & number of explanations for each recommendation \\
		\hline
		$\hat{r}_{u, i}$ & score predicted for user $u$ on item $i$ \\
		$\hat{r}_{u, i, e}$ & score predicted for user $u$ on explanation $e$ of item $i$ \\
		\hline
	\end{tabular}
	\label{tbl:notation}
\end{table}

\section{Problem Formulation} \label{sec:problem}

\begin{figure}
	\centering
	\includegraphics[scale=0.5]{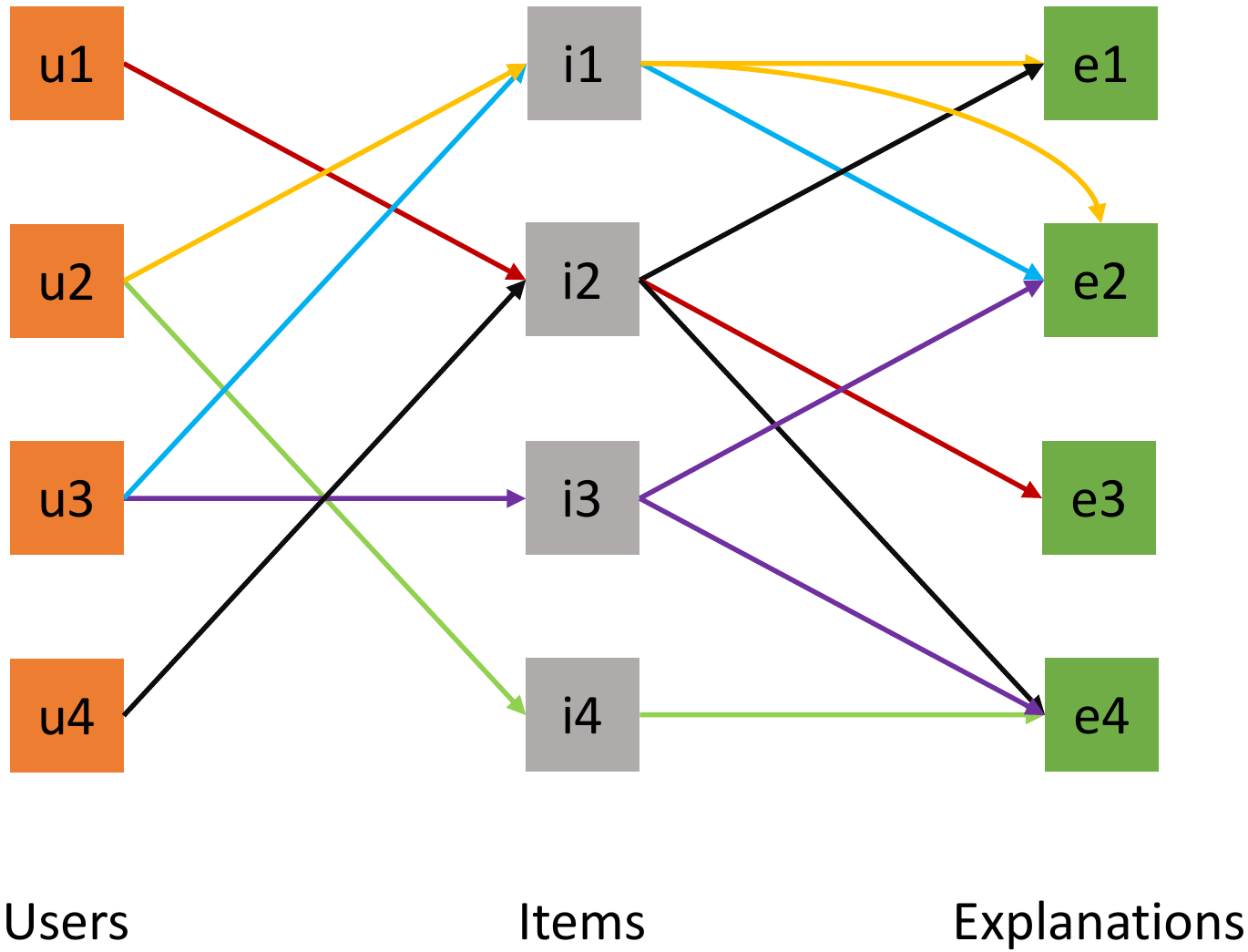}
	\caption{Illustration of user-item-explanation interactions.}
	\label{fig:interaction}
\end{figure}

The key notations and concepts for the problems are presented in Table \ref{tbl:notation}.
We use $\mathcal{U}$ to denote the set of all users, $\mathcal{I}$ the set of all items and $\mathcal{E}$ the set of all explanations.
Then the historical interaction set is given by $\mathcal{T} \subseteq \mathcal{U} \times \mathcal{I} \times \mathcal{E}$ (an illustrating example of such interaction is depicted in Fig. \ref{fig:interaction}).
In the following, we first introduce item ranking and explanation ranking respectively, and then the item-explanation joint-ranking.

\subsection{Item Ranking}

Personalized recommendation aims at providing a user with a ranked list of items that he/she never interacted with before.
For each user $u \in \mathcal{U}$, the list of $M$ items can be generated as follows,
\begin{equation}
	\text{Top}(u, M) := \mathop{\arg\max}_{i \in \mathcal{I} / \mathcal{I}_u}^{M} \hat{r}_{u, \underline{i}}
	\label{eqn:topi}
\end{equation}
where $\hat{r}_{u, \underline{i}}$ is the predicted score for a user $u$ on item $i$, and $\mathcal{I} / \mathcal{I}_u$ denotes the set of items on which user $u$ has no interactions. In Eq. \eqref{eqn:topi}, $i$ is underlined, which means that we aim to rank the items.

\subsection{Explanation Ranking}

Explanation ranking is the task of finding a list of appropriate explanations for a user-item pair to justify the recommendation.
Formally, given a user $u \in \mathcal{U}$ and an item $i \in \mathcal{I}$, the goal of this task is to rank the entire collection of explanations $\mathcal{E}$, and select the top $N$ to reason why the item $i$ is recommended.
Specifically, we define this list of top $N$ explanations as:
\begin{equation}
	\text{Top}(u, i, N) := \mathop{\arg\max}_{e \in \mathcal{E}}^{N} \hat{r}_{u, i, \underline{e}}
	\label{eqn:topn}
\end{equation}
where $\hat{r}_{u, i, \underline{e}}$ is the estimated score of explanation $e$ for a given user-item pair $(u, i)$, which could be given by a recommendation model or by the user's true behavior.

\subsection{Item-Explanation Joint-Ranking}

The preceding tasks solely rank either items or explanations.
In this task, we further investigate whether it is possible to find an ideal item-explanation pair for a user, to whom the explanation best justifies the item that he/she likes the most.
To this end, we treat each pair of item-explanation as a joint unit, and then rank these units.
Specifically, for each user $u \in \mathcal{U}$, a ranked list of $M$ item-explanation pairs can be produced as follows,
\begin{equation}
	\text{Top}(u, M) := \mathop{\arg\max}_{i \in \mathcal{I} / \mathcal{I}_u, e \in \mathcal{E}}^{M} \hat{r}_{u, \underline{i, e}}
	\label{eqn:topn2}
\end{equation}
where $\hat{r}_{u, \underline{i, e}}$ is the predicted score for a given user $u$ on the item-explanation pair ($i$, $e$).

We see that either item ranking task or explanation ranking task is a special case of this item-explanation joint-ranking task.
Concretely, Eq. \eqref{eqn:topn2} degenerates to Eq. \eqref{eqn:topi} when explanation $e$ is fixed, while it reduces to Eq. \eqref{eqn:topn} if item $i$ is already known.

\section{Our Framework for Ranking Tasks} \label{sec:model}

\subsection{Joint-Ranking Reformulation}

Suppose we have an ideal model that can perform the aforementioned joint-ranking task.
During the prediction stage as in Eq. \eqref{eqn:topn2}, there would be $\left| \mathcal{I} \right| \times \left| \mathcal{E} \right|$ candidate item-explanation pairs to rank for each user $u \in \mathcal{U}$.
The runtime complexity is then $O (\left| \mathcal{U} \right| \cdot \left| \mathcal{I} \right| \cdot \left| \mathcal{E} \right|)$, which makes this task impractical, compared with the traditional recommendation task's $O (\left| \mathcal{U} \right| \cdot \left| \mathcal{I} \right|)$ complexity.

To reduce the complexity, we reformulate the joint-ranking task by performing ranking for items and explanations simultaneously but separately.
In this way, we are also able to investigate the relationship between item ranking and explanation ranking, e.g., improving the performance of both.
Specifically, during the testing stage, we first follow Eq. \eqref{eqn:topi} to rank items for each user $u \in \mathcal{U}$, which has the runtime complexity of $O (\left| \mathcal{U} \right| \cdot \left| \mathcal{I} \right|)$.
After that, for $M$ recommendations for each user, we can rank and select explanations to justify each of them according to Eq. \eqref{eqn:topn}.
The second step's complexity is $O (\left| \mathcal{U} \right| \cdot M \cdot \left| \mathcal{E} \right|)$, but since $M$ is a constant and $\left| \mathcal{E} \right| \ll \left| \mathcal{I} \right|$ (see Table \ref{tbl:dataset}), the overall complexity of the two steps is $O (\left| \mathcal{U} \right| \cdot \left| \mathcal{I} \right|)$.

In the following, we first analyze the drawback of a conventional Tensor Factorization (TF) model when being applied to the explanation ranking problem, and then introduce our solution BPER.
Second, we show how to further enhance BPER by utilizing the semantic features of textual explanations (denoted as BPER+).
Third, we illustrate their relation to two typical TF methods CD and PITF.
At last, we integrate the explanation ranking with item ranking into a multi-task learning framework as a joint-ranking task.

\subsection{Bayesian Personalized Explanation Ranking (BPER)} \label{sec:bper}

To perform explanation ranking, the score $\hat{r}_{u, i, e}$ on each explanation $e \in \mathcal{E}$ for a given user-item pair $(u, i)$ must be estimated.
As the user-item-explanation ternary relations $\mathcal{T} = \{(u, i, e) | u \in \mathcal{U}, i \in \mathcal{I}, e \in \mathcal{E}\}$ form an interaction cube, we are inspired to employ factorization models to predict this type of scores.
There are a number of tensor factorization techniques \cite{bhargava2015and, ioannidis2019coupled}, such as Tucker Decomposition (TD) \cite{Springer66-TD}, Canonical Decomposition (CD) \cite{Springer70-CD} and High Order Singular Value Decomposition (HOSVD) \cite{SIAM00-HOSVD}.
Intuitively, one would adopt CD, because of its linear runtime complexity in terms of both training and prediction \cite{WSDM10-PITF} and its close relation to Matrix Factorization (MF) \cite{NIPS08-PMF} that has been extensively studied in recent years for item recommendation.
Formally, according to CD, the score $\hat{r}_{u, i, e}$ of user $u$ on item $i$'s explanation $e$ can be estimated by the sum over the element-wise multiplication of the user's latent factors $\mathbf{p}_u$, the item's $\mathbf{q}_i$ and the explanation's $\mathbf{o}_e$:
\begin{equation}
	\hat{r}_{u, i, e} = (\mathbf{p}_u \odot \mathbf{q}_i)^\top \mathbf{o}_e = \sum_{k = 1}^d p_{u, k} \cdot q_{i, k} \cdot o_{e, k}
	\label{eqn:cd}
\end{equation}
where $\odot$ denotes the element-wise multiplication of two vectors.

However, this method may not be effective enough due to the inherent sparsity problem of the ternary data as we discussed before.
Since each user-item pair $(u, i)$ in the training set $\mathcal{T}$ is unlikely to have interactions with many explanations in $\mathcal{E}$, the data sparsity problem for explanation ranking is severer than that for item recommendation.
Simply multiplying the three vectors would hurt the performance of explanation ranking, which is evidenced by our experimental results in Section \ref{sec:results}.

To mitigate such an issue and to improve the effectiveness of explanation ranking, we propose to separately estimate the user $u$'s preference score $\hat{r}_{u, e}$ on explanation $e$ and the item $i$'s appropriateness score $\hat{r}_{i, e}$ for explanation $e$.
To this end, we perform two sets of matrix factorization, rather than employing one single TF model.
In this way, the sparsity problem would be considerably alleviated, since the data are reduced to two collections of binary relations, both of which are similar to the case of item recommendation discussed above.
At last, the two scores $\hat{r}_{u, e}$ and $\hat{r}_{i, e}$ are combined linearly through a hyper-parameter $\mu$.
Specifically, the score of user $u$ for item $i$ on explanation $e$ is predicted as follows,
\begin{equation}
	\left\{
	\begin{array}{l}
		\hat{r}_{u, e} = \mathbf{p}_u^\top \mathbf{o}_e^U + b_e^U = \sum_{k = 1}^d p_{u, k} \cdot o_{e, k}^U + b_e^U \\
		\hat{r}_{i, e} = \mathbf{q}_i^\top \mathbf{o}_e^I + b_e^I = \sum_{k = 1}^d q_{i, k} \cdot o_{e, k}^I + b_e^I \\
		\hat{r}_{u, i, e} = \mu \cdot \hat{r}_{u, e} + (1 - \mu) \cdot \hat{r}_{i, e}
	\end{array}
	\right.
	\label{eqn:predict}
\end{equation}
where $\{\mathbf{o}_e^U, b_e^U\}$ and $\{\mathbf{o}_e^I, b_e^I\}$ are two different sets of latent factors for explanations, corresponding to users and items respectively.

Since selecting explanations that are likely to be perceived helpful by users is inherently a ranking-oriented task, directly modeling the relative order of explanations is thus more effective than simply predicting their absolute scores.
The Bayesian Personalized Ranking (BPR) criterion \cite{UAI09-BPR} meets such an optimization requirement.
Intuitively, a user would be more likely to appreciate explanations that cater to her own preferences, while those that do not fit one's interests would be less attractive to the user.
Similarly, some explanations might be more suitable to describe certain items, while other explanations might not.
To build such type of pair-wise preferences, we use the first two rows in Eq. \eqref{eqn:predict} to compute the difference between two explanations for both user $u$ and item $i$ as follows,
\begin{equation}
	\left\{
	\begin{array}{l}
		\hat{r}_{u, ee'} = \hat{r}_{u, e} -  \hat{r}_{u, e'} \\
		\hat{r}_{i, ee''} = \hat{r}_{i, e} - \hat{r}_{i, e''}
	\end{array}
	\right.
	\label{eqn:diff}
\end{equation}
which respectively reflect user $u$'s interest in explanation $e$ over $e'$, and item $i$'s appropriateness for explanation $e$ over $e''$.

With the scores $\hat{r}_{u, ee'}$ and $\hat{r}_{u, ee''}$, we can then adopt the BPR criterion \cite{UAI09-BPR} to minimize the following objective function:
\begin{equation}
		\min_{\Theta} \sum_{u \in \mathcal{U}} \sum_{i \in \mathcal{I}_u} \sum_{e \in \mathcal{E}_{u, i}} \Big[ \sum_{e' \in \mathcal{E} / \mathcal{E}_u} - \ln \sigma (\hat{r}_{u, ee'}) + \sum_{e'' \in \mathcal{E} / \mathcal{E}_i} - \ln \sigma (\hat{r}_{i, ee''}) \Big] + \lambda \left| \left| \Theta \right| \right|_F^2
	\label{eqn:exp}
\end{equation}
where $\sigma(\cdot)$ denotes the sigmoid function, $\mathcal{I}_{u}$ represents the set of items that user $u$ has interacted with, $\mathcal{E}_{u, i}$ is the set of explanations in the training set for the user-item pair $(u, i)$, $\mathcal{E} / \mathcal{E}_u$ and $\mathcal{E} / \mathcal{E}_i$ respectively correspond to explanations that user $u$ and item $i$ have not interacted with, $\Theta$ is the model parameter, and $\lambda$ is the regularization coefficient.

From Eq. \eqref{eqn:exp}, we can see that there are two explanation tasks to be learned respectively, corresponding to users and items.
During the training stage, we allow them to be equally important, since we have a hyper-parameter $\mu$ in Eq. \eqref{eqn:predict} to balance their importance during the testing stage.
The effect of this parameter is studied in Section \ref{sec:mu}.
After the model parameters are estimated, we can rank explanations according to Eq. \eqref{eqn:topn} for each user-item pair in the testing set.
As we model the explanation ranking task under BPR criterion, we accordingly name our method Bayesian Personalized Explanation Ranking (BPER).
To learn the model parameter $\Theta$, we draw on the widely used stochastic gradient descent algorithm to optimize the objective function in Eq. \eqref{eqn:exp}.
Specifically, we first randomly initialize the parameters, and then repeatedly update them by uniformly taking samples from the training set and computing the gradients w.r.t. the parameters, until the convergence of the algorithm.
The complete learning steps are shown in Algorithm \ref{alg:bper}.

\subsection{BERT-enhanced BPER (BPER+)}

The BPER model only exploits the IDs of users, items and explanations to infer their relation for explanation ranking.
However, this makes the rich semantic features of the explanations, which could also capture the relation between explanations, under-explored.
For example, ``\textit{the acting is good}'' and ``\textit{the acting is great}'' for movie recommendation both convey a positive sentiment with a similar meaning, so their ranks are expected to be close.
Hence, we further investigate whether such features could help to enhance BPER.
As a feature extractor, we opt for BERT \cite{NAACL19-BERT}, a well-known pre-trained language model, whose effectiveness has been demonstrated on a wide range of natural language understanding tasks.
Specifically, we first add a special [CLS] token at the beginning of a textual explanation $e$, e.g., ``\textit{[CLS] the acting is great}''.
After passing it through BERT, we can obtain the aggregate representation (corresponding to [CLS]) that encodes the explanation's overall semantics.
To match the dimension of latent factors in our model, we apply a linear layer to this vector, resulting in $\mathbf{o}_e^{BERT}$.
Then, we enhance the two ID-based explanation vectors $\mathbf{o}_e^U$ and $\mathbf{o}_e^I$ in Eq. \eqref{eqn:predict} by multiplying $\mathbf{o}_e^{BERT}$, resulting in $\mathbf{o}_e^{U+}$ and $\mathbf{o}_e^{I+}$.
\begin{equation}
	\left\{
	\begin{array}{l}
		\mathbf{o}_e^{U+} = \mathbf{o}_e^U \odot \mathbf{o}_e^{BERT} \\
		\mathbf{o}_e^{I+} = \mathbf{o}_e^I \odot \mathbf{o}_e^{BERT}
	\end{array}
	\right.
	\label{eqn:bperp}
\end{equation}

\begin{algorithm}[H]
	\caption{Bayesian Personalized Explanation Ranking (BPER)}
	\label{alg:bper}
	\begin{algorithmic}[1]
		\REQUIRE training set $\mathcal{T}$, dimension of latent factors $d$, learning rate $\gamma$, regularization coefficient $\lambda$, iteration number $T$
		\ENSURE model parameters $\Theta = \{\mathbf{P}, \mathbf{Q}, \mathbf{O}^U, \mathbf{O}^I, \mathbf{b}^U, \mathbf{b}^I\}$
		\STATE Initialize $\Theta$, including $\mathbf{P} \gets \mathbb{R}^{\left| \mathcal{U} \right| \times d}$, $\mathbf{Q} \gets \mathbb{R}^{\left| \mathcal{I} \right| \times d}$, $\mathbf{O}^U \gets \mathbb{R}^{\left| \mathcal{E} \right| \times d}, \mathbf{O}^I \gets \mathbb{R}^{\left| \mathcal{E} \right| \times d}, \mathbf{b}^U \gets \mathbb{R}^{\left| \mathcal{E} \right|}, \mathbf{b}^I \gets \mathbb{R}^{\left| \mathcal{E} \right|}$
		\FOR{$t_1 = 1$ to $T$}
		\FOR{$t_2 = 1$ to $\left| \mathcal{T} \right|$}
		\STATE Uniformly draw $(u, i, e)$ from $\mathcal{T}$, $e'$ from $\mathcal{E} / \mathcal{E}_u$, and $e''$ from $\mathcal{E} / \mathcal{E}_i$
		\STATE $\hat{r}_{u, ee'} \gets \hat{r}_{u, e} -  \hat{r}_{u, e'}$, $\hat{r}_{i, ee''} \gets \hat{r}_{i, e} - \hat{r}_{i, e''}$
		\STATE $x \gets - \sigma (-\hat{r}_{u, ee'})$, $y \gets - \sigma (-\hat{r}_{i, ee''})$
		\STATE $\mathbf{p}_u \gets \mathbf{p}_u - \gamma \cdot (x \cdot (\mathbf{o}_e^U - \mathbf{o}_{e'}^U) + \lambda \cdot \mathbf{p}_u)$
		\STATE $\mathbf{q}_i \gets \mathbf{q}_i - \gamma \cdot (y \cdot (\mathbf{o}_e^I - \mathbf{o}_{e''}^I) + \lambda \cdot \mathbf{q}_i)$
		\STATE $\mathbf{o}_e^U \gets \mathbf{o}_e^U - \gamma \cdot (x \cdot \mathbf{p}_u + \lambda \cdot \mathbf{o}_e^U)$
		\STATE $\mathbf{o}_{e'}^U \gets \mathbf{o}_{e'}^U - \gamma \cdot (- x \cdot \mathbf{p}_u + \lambda \cdot \mathbf{o}_{e'}^U)$
		\STATE $\mathbf{o}_e^I \gets \mathbf{o}_e^I - \gamma \cdot (y \cdot \mathbf{q}_i + \lambda \cdot \mathbf{o}_e^I)$
		\STATE $\mathbf{o}_{e''}^I \gets \mathbf{o}_{e''}^I - \gamma \cdot (- y \cdot \mathbf{q}_i + \lambda \cdot \mathbf{o}_{e''}^I)$
		\STATE $b_e^U \gets b_e^U - \gamma \cdot (x + \lambda \cdot b_e^U)$
		\STATE $b_{e'}^U \gets b_{e'}^U - \gamma \cdot (- x + \lambda \cdot b_{e'}^U)$
		\STATE $b_e^I \gets b_e^I - \gamma \cdot (y + \lambda \cdot b_e^I)$
		\STATE $b_{e''}^I \gets b_{e''}^I - \gamma \cdot (- y + \lambda \cdot b_{e''}^I)$
		\ENDFOR
		\ENDFOR
	\end{algorithmic}
\end{algorithm}

To predict the score for $(u, i, e)$ triplet, we replace $\mathbf{o}_e^U$ and $\mathbf{o}_e^I$ in Eq. \eqref{eqn:predict} with $\mathbf{o}_e^{U+}$ and $\mathbf{o}_e^{I+}$.
Then we use Eq. \eqref{eqn:exp} as the objective function, which can be optimized via back-propagation.
In Eq. \eqref{eqn:bperp}, we adopt the multiplication operation simply to verify the feasibility of incorporating semantic features.
The model may be further improved by more sophisticated operations, e.g., multi-layer perceptron (MLP), but we leave the exploration for future work.

Notice that, BPER is a general method that only requires the IDs of users, items and explanations, which makes it very flexible when being adapted to other explanation styles (e.g., images \cite{SIGIR19-VECF}).
However, it may suffer from the common cold-start issue as with other recommender systems.
BPER+ could mitigate this issue to some extent, because besides IDs it also considers the semantic relation between textual explanations via BERT, which can connect new explanations with existing ones.
As the first work on ranking explanations for recommendations, we opt to make both methods relatively simple for reproducibility purpose.
In this way, it is also easy to observe the experimental results (such as the impact of explanation task on recommendation task), without the interference of other factors.

\begin{figure*}
	\centering
	\begin{minipage}{0.5\textwidth}
		\begin{flushright}
			\subfigure[Our Bayesian Personalized Explanation Ranking (BPER)]{\includegraphics[scale=0.34]{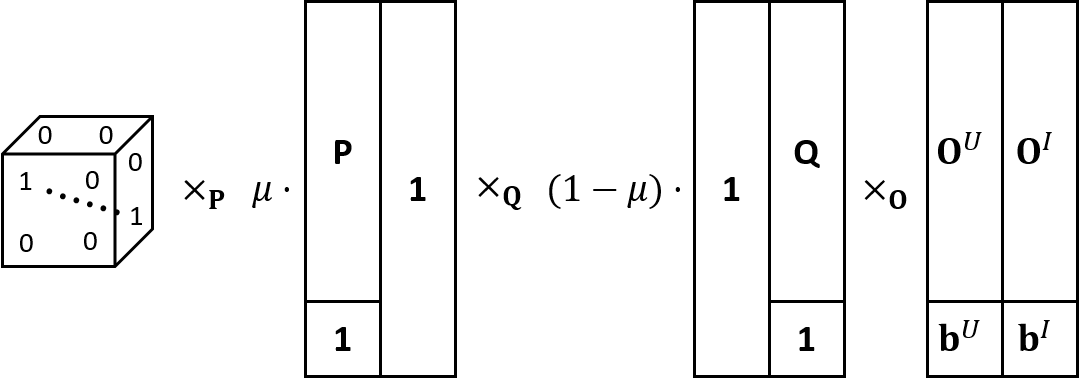}}
			\subfigure[Our BERT-enhanced BPER (BPER+)]{\includegraphics[scale=0.34]{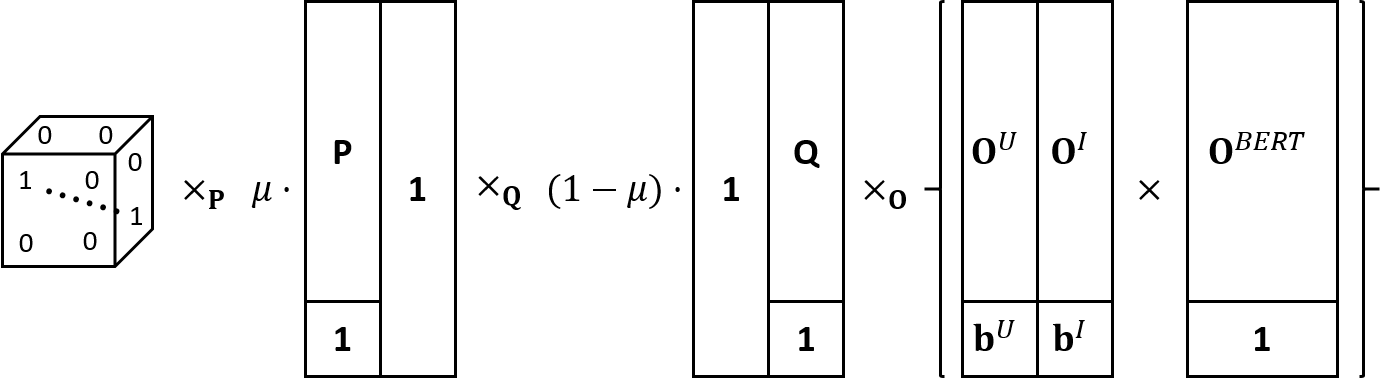}}
		\end{flushright}
	\end{minipage}
	\hspace{15mm}
	\begin{minipage}{0.37\textwidth}
		\subfigure[Canonical Decomposition (CD)]{\includegraphics[scale=0.34]{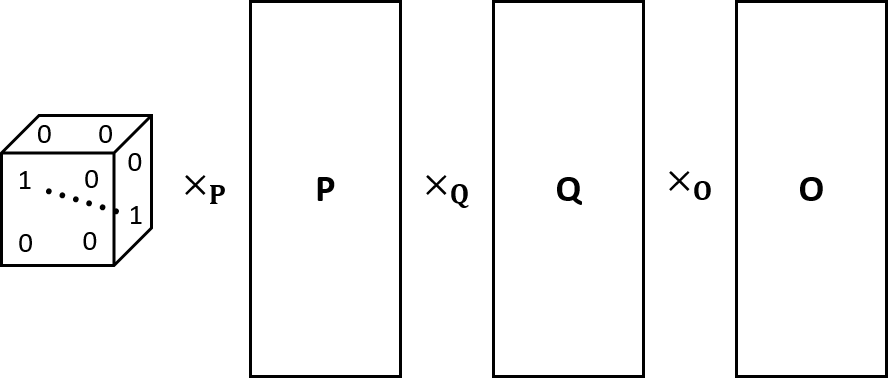}}
		\subfigure[Pairwise Interaction Tensor Factorization (PITF)]{\includegraphics[scale=0.34]{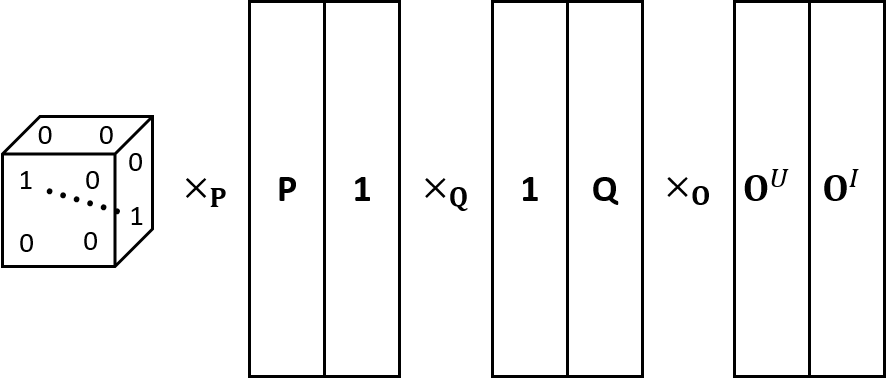}}
	\end{minipage}
	\caption{Tensor Factorization models. The three matrices (i.e., $\mathbf{P}$, $\mathbf{Q}$, $\mathbf{O}$) are model parameters. Our BPER and BPER+ can be regarded as special cases of CD, while PITF can be seen as a special case of our BPER and BPER+.}
	\label{fig:model}
\end{figure*}

\subsection{Relation between BPER, BPER+, CD, and PITF}

In fact, our Bayesian Personalized Explanation Ranking (BPER) model is a type of Tensor Factorization (TF), so we analyze its relation to two closely related TF methods: Canonical Decomposition (CD) \cite{Springer70-CD} and Pairwise Interaction Tensor Factorization (PITF) \cite{WSDM10-PITF}.
On the one hand, in theory BPER can be considered as a special case of the CD model.
Suppose the dimensionality of BPER is $2 \cdot d + 2$, we can reformulate it as CD in the following,
\begin{equation}
	\begin{aligned}
		p_{u, k}^{CD} &=
		\begin{cases}
			\mu \cdot p_{u, k}, & \mbox{if } k \leq d \\
			\mu, & \mbox{else}
		\end{cases} \\
		q_{i, k}^{CD} &=
		\begin{cases}
			(1 - \mu) \cdot q_{i, k}, & \mbox{if } k > d \mbox{ and } k \leq 2 \cdot d  \\
			1 - \mu, & \mbox{else}
		\end{cases} \\
		o_{e, k}^{CD} &=
		\begin{cases}
			o_{e, k}^U, & \mbox{if } k \leq d \\
			o_{e, k}^I, & \mbox{else if } k \leq 2 \cdot d \\
			b_e^U, & \mbox{else if } k = 2 \cdot d + 1 \\
			b_e^I, & \mbox{else}
		\end{cases}
	\end{aligned}
	\label{eqn:relation}
\end{equation}
where the parameter $\mu$ is a constant.

On the other hand, PITF can be seen as a special case of our BPER.
Formally, its predicted score $\hat{r}_{u, i, e}$ for the user-item-explanation triplet $(u, i, e)$ can be calculated by:
\begin{equation}
	\hat{r}_{u, i, e} = \mathbf{p}_u^\top \mathbf{o}_e^U + \mathbf{q}_i^\top \mathbf{o}_e^I = \sum_{k = 1}^d p_{u, k} \cdot o_{e, k}^U + \sum_{k = 1}^d q_{i, k} \cdot o_{e, k}^I
	\label{eqn:pitf}
\end{equation}

We can see that our BPER degenerates to PITF if in Eq. \eqref{eqn:predict} we remove the bias terms $b_e^U$ and $b_e^I$ and set the hyper-parameter $\mu$ to 0.5, which means that the two types of scores for users and items are equally important to the explanation ranking task.

Although CD is more general than our BPER, its performance may be affected by the data sparsity issue as discussed before.
Our BPER could mitigate this problem given its explicitly designed structure that may be difficult for CD to learn from scratch.
When comparing with PITF, we can find that the parameter $\mu$ in BPER is able to balance the importance of the two types of scores, corresponding to users and items, which makes our BPER more expressive than PITF and hence likely reach better ranking quality.

In a similar way, BPER+ can also be rewritten as CD or PITF.
Concretely, by revising the last part of Eq. \eqref{eqn:relation} as the following formula, BPER+ can be seen as CD.
When $\mathbf{o}_e^{BERT} = [1, ..., 1]^\top$, BPER+ is equal to BPER, so it can be easily converted into PITF.
The graphical illustration of the four models is shown in Fig. \ref{fig:model}.

\begin{equation}
	\begin{aligned}
		o_{e, k}^{CD} &=
		\begin{cases}
			o_{e, k}^U \cdot o_{e, k}^{BERT}, & \mbox{if } k \leq d \\
			o_{e, k}^I \cdot o_{e, k}^{BERT}, & \mbox{else if } k \leq 2 \cdot d \\
			b_e^U, & \mbox{else if } k = 2 \cdot d + 1 \\
			b_e^I, & \mbox{else}
		\end{cases}
	\end{aligned}
\end{equation}

\subsection{Joint-Ranking on BPER (BPER-J)}

Owing to BPER's flexibility to accommodate various explanation styles as discussed before, we perform the joint-ranking on it.
Specifically, we incorporate the two tasks of explanation ranking and item recommendation into a unified multi-task learning framework, so as to find a good solution that benefits both of them.

For recommendation, we adopt Singular Value Decomposition (SVD) model \cite{Computer09-MF} to predict the score $\hat{r}_{u, i}$ of user $u$ on item $i$:
\begin{equation}
	\hat{r}_{u, i} = \mathbf{p}_u^\top \mathbf{q}_i + b_i = \sum_{k = 1}^d p_{u, k} \cdot q_{i, k} + b_i
\end{equation}
where $b_i$ is the bias term for item $i$.
Notice that, the latent factors $\mathbf{p}_u$ and $\mathbf{q}_i$ are shared with those for explanation ranking in Eq. \eqref{eqn:predict}.
In essence, item recommendation is also a ranking task that can be optimized using BPR criteria \cite{UAI09-BPR}, so we first compute the preference difference $\hat{r}_{u, ii'}$ between a pair of items $i$ and $i'$ to a user $u$ as follows,
\begin{equation}
	\hat{r}_{u, ii'} = \hat{r}_{u, i} - \hat{r}_{u, i'}
\end{equation}
which can then be combined with the task of explanation ranking in Eq. \eqref{eqn:exp} to form the following objective function for joint-ranking:
\begin{equation}
	\begin{split}
		\min_{\Theta} \sum_{u \in \mathcal{U}} \sum_{i \in \mathcal{I}_u} \Big[\sum_{i' \in \mathcal{I} / \mathcal{I}_u} - \ln \sigma (\hat{r}_{u, ii'}) + \alpha \sum_{e \in \mathcal{E}_{u, i}} \Big(\sum_{e' \in \mathcal{E} / \mathcal{E}_u} \\ - \ln \sigma (\hat{r}_{u, ee'}) + \sum_{e'' \in \mathcal{E} / \mathcal{E}_i} - \ln \sigma (\hat{r}_{i, ee''})\Big)\Big] + \lambda \left| \left| \Theta \right| \right|_F^2
	\end{split}
\end{equation}
where the parameter $\alpha$ can be fine-tuned to balance the learning of the two tasks.

We name this method \textbf{BPER-J} where J stands for joint-ranking.
Similar to BPER, we can update each parameter of BPER-J via stochastic gradient descent (see Algorithm \ref{alg:bper-j}).

\section{Experimental Setup} \label{sec:setup}

\subsection{Datasets}

To compare the ranking performance of different methods, it is expected that the datasets contain user-item-explanation interaction triplets.
The datasets could be manually constructed as in \cite{WWW20-Office}, but we are not given access to such datasets.
Therefore, we adopt three public datasets\footnote{https://github.com/lileipisces/EXTRA} \cite{SIGIR21-EXTRA}, where the explanations are automatically extracted from user reviews via near-duplicate detection, which ensures that the explanations are commonly used by users.
Specifically, the datasets are from different domains, including Amazon Movies \& TV\footnote{http://jmcauley.ucsd.edu/data/amazon}, TripAdvisor\footnote{https://www.tripadvisor.com} for hotels and Yelp\footnote{https://www.yelp.com/dataset/challenge} for restaurants.
Each record in the three datasets consists of user ID, item ID, and one or multiple explanation IDs, and thus results in one or multiple user-item-explanation triplets.
Moreover, each explanation ID appears no less than 5 times.
The statistics of the three datasets are presented in Table \ref{tbl:dataset}.
As it can be seen, the data sparsity issue on the three datasets is very severe.

\begin{algorithm}[H]
	\caption{Joint-Ranking on BPER (BPER-J)}
	\label{alg:bper-j}
	\begin{algorithmic}[1]
		\REQUIRE training set $\mathcal{T}$, dimension of latent factors $d$, learning rate $\gamma$, regularization coefficients $\alpha$ and $\lambda$, iteration number $T$
		\ENSURE model parameters $\Theta = \{\mathbf{P}, \mathbf{Q}, \mathbf{O}^U, \mathbf{O}^I, \mathbf{b}, \mathbf{b}^U, \mathbf{b}^I\}$
		\STATE Initialize $\Theta$, including $\mathbf{P} \gets \mathbb{R}^{\left| \mathcal{U} \right| \times d}$, $\mathbf{Q} \gets \mathbb{R}^{\left| \mathcal{I} \right| \times d}$, $\mathbf{O}^U \gets \mathbb{R}^{\left| \mathcal{E} \right| \times d}, \mathbf{O}^I \gets \mathbb{R}^{\left| \mathcal{E} \right| \times d}, \mathbf{b} \gets \mathbb{R}^{\left| \mathcal{I} \right|}, \mathbf{b}^U \gets \mathbb{R}^{\left| \mathcal{E} \right|}, \mathbf{b}^I \gets \mathbb{R}^{\left| \mathcal{E} \right|}$
		\FOR{$t_1 = 1$ to $T$}
		\FOR{$t_2 = 1$ to $\left| \mathcal{T} \right|$}
		\STATE Uniformly draw $(u, i, e)$ from $\mathcal{T}$, $e'$ from $\mathcal{E} / \mathcal{E}_u$, $e''$ from $\mathcal{E} / \mathcal{E}_i$, and $i'$ from $\mathcal{I} / \mathcal{I}_u$
		\STATE $\hat{r}_{u, ee'} \gets \hat{r}_{u, e} -  \hat{r}_{u, e'}$, $\hat{r}_{i, ee''} \gets \hat{r}_{i, e} - \hat{r}_{i, e''}$, $\hat{r}_{u, ii'} \gets \hat{r}_{u, i} - \hat{r}_{u, i'}$
		\STATE $x \gets - \alpha \cdot \sigma (-\hat{r}_{u, ee'})$, $y \gets - \alpha \cdot \sigma (-\hat{r}_{i, ee''})$, $z \gets - \sigma (-\hat{r}_{u, ii'})$
		\STATE $\mathbf{p}_u \gets \mathbf{p}_u - \gamma \cdot (x \cdot (\mathbf{o}_e^U - \mathbf{o}_{e'}^U) + z \cdot (\mathbf{q}_i - \mathbf{q}_{i'}) + \lambda \cdot \mathbf{p}_u)$
		\STATE $\mathbf{q}_i \gets \mathbf{q}_i - \gamma \cdot (y \cdot (\mathbf{o}_e^I - \mathbf{o}_{e''}^I) + z \cdot \mathbf{p}_u + \lambda \cdot \mathbf{q}_i)$
		\STATE $\mathbf{q}_{i'} \gets \mathbf{q}_{i'} - \gamma \cdot (-z \cdot \mathbf{p}_u + \lambda \cdot \mathbf{q}_{i'})$
		\STATE $\mathbf{o}_e^U \gets \mathbf{o}_e^U - \gamma \cdot (x \cdot \mathbf{p}_u + \lambda \cdot \mathbf{o}_e^U)$
		\STATE $\mathbf{o}_{e'}^U \gets \mathbf{o}_{e'}^U - \gamma \cdot (- x \cdot \mathbf{p}_u + \lambda \cdot \mathbf{o}_{e'}^U)$
		\STATE $\mathbf{o}_e^I \gets \mathbf{o}_e^I - \gamma \cdot (y \cdot \mathbf{q}_i + \lambda \cdot \mathbf{o}_e^I)$
		\STATE $\mathbf{o}_{e''}^I \gets \mathbf{o}_{e''}^I - \gamma \cdot (- y \cdot \mathbf{q}_i + \lambda \cdot \mathbf{o}_{e''}^I)$
		\STATE $b_i \gets b_i - \gamma \cdot (z + \lambda \cdot b_i)$
		\STATE $b_{i'} \gets b_{i'} - \gamma \cdot (-z + \lambda \cdot b_{i'})$
		\STATE $b_e^U \gets b_e^U - \gamma \cdot (x + \lambda \cdot b_e^U)$
		\STATE $b_{e'}^U \gets b_{e'}^U - \gamma \cdot (- x + \lambda \cdot b_{e'}^U)$
		\STATE $b_e^I \gets b_e^I - \gamma \cdot (y + \lambda \cdot b_e^I)$
		\STATE $b_{e''}^I \gets b_{e''}^I - \gamma \cdot (- y + \lambda \cdot b_{e''}^I)$
		\ENDFOR
		\ENDFOR
	\end{algorithmic}
\end{algorithm}

\begin{table}[!tbp]
	\centering
	\caption{Statistics of the datasets. Density is \#triplets divided by \#users $\times$ \#items $\times$ \#explanations.}
	\begin{tabular}{l|r|r|r}
		\hline
		& \textbf{Amazon Movies \& TV} & \textbf{TripAdvisor} & \textbf{Yelp} \\
		\hline
		\# of users                & 109,121                & 123,374               & 895,729 \\
		\# of items                 & 47,113                & 200,475               & 164,779 \\
		\# of explanations               & 33,767              & 76,293            & 126,696 \\
		\# of $(u, i)$ pairs & 569,838 & 1,377,605 & 2,608,860 \\
		\# of $(u, i, e)$ triplets    & 793,481                & 2,618,340               & 3,875,118 \\
		\# of explanations / $(u, i)$ pair & 1.39 & 1.90 & 1.49 \\
		Density ($\times 10^{-10}$)  & 		45.71   	&  	13.88			  & 2.07 \\
		\hline
	\end{tabular}
	\label{tbl:dataset}
\end{table}

Table \ref{tbl:case} shows 5 example explanations taken from the three datasets.
As we can see, all the explanations are quite concise and informative, which could prevent from overwhelming users, a critical issue for explainable recommendation \cite{CSCW00-CF}.
Also, short explanations can be mobile-friendly, since it is difficult for a small screen to fit much content.
Moreover, the explanations from different datasets well suit the target application domains, such as ``\textit{a wonderful movie for all ages}'' for movies and ``\textit{comfortable hotel with good facilities}'' for hotels.
Explanations with negative sentiment can also be observed, e.g., ``\textit{the place is awful}'', which can be used to justify why some items are dis-recommended \cite{SIGIR14-EFM}.
Hence, we believe that the datasets are very suitable for our explanation ranking experiment.

\subsection{Compared Methods}

To evaluate the performance of explanation ranking task, where the user-item pairs are given, we adopt the following baselines.
Notice that, we omit the comparison with Tucker Decomposition (TD) \cite{Springer66-TD}, because it takes cubic time to run and we also find that it does not perform better than CD in our trial experiment.
\begin{itemize}
	\item \textbf{RAND}: It is a weak baseline that randomly picks up explanations from the explanation collection $\mathcal{E}$.
	It is devised to examine whether personalization is needed for explanation ranking.
	\item \textbf{RUCF}: Revised User-based Collaborative Filtering.
	Because traditional CF methods \cite{CSCW94-UCF, WWW01-ICF} cannot be directly applied to the ternary data, we make some modifications to their formula, following \cite{PKDD07-FolkRank}.
	The similarity between two users is measured by their associated explanation sets via Jaccard Index.
	When predicting a score for the $(u, i, e)$ triplet, we first find users associated with the same item $i$ and explanation $e$, i.e., $\mathcal{U}_i \cap \mathcal{U}_e$, from which we then find the ones appearing in user $u$'s neighbor set $\mathcal{N}_u$.
	\begin{equation}
		\hat{r}_{u, i, e} = \sum_{u' \in \mathcal{N}_u \cap (\mathcal{U}_i \cap \mathcal{U}_e)} s_{u, u'}
		\mbox{ where }
		s_{u, u'} = \frac{\vert \mathcal{E}_u \cap \mathcal{E}_{u'} \vert}{\vert \mathcal{E}_u \cup \mathcal{E}_{u'} \vert}
		\label{eqn:icf}
	\end{equation}
	\item \textbf{RICF}: Revised Item-based Collaborative Filtering.
	This method predicts a score for a triplet from the perspective of items, whose formula is similar to Eq. \eqref{eqn:icf}.
	\item \textbf{CD}: Canonical Decomposition \cite{Springer70-CD} as shown in Eq. \eqref{eqn:cd}.
	This method only predicts one score instead of two for the triplet $(u, i, e)$, so its objective function shown below is slightly different from ours in Eq. \eqref{eqn:exp}.
	\begin{equation}
		\min_{\Theta} \sum_{u \in \mathcal{U}} \sum_{i \in \mathcal{I}_u} \sum_{e \in \mathcal{E}_{u, i}} \sum_{e' \in \mathcal{E} / \mathcal{E}_{u, i}} - \ln \sigma (\hat{r}_{u, i, ee'}) + \lambda \left| \left| \Theta \right| \right|_F^2
		\label{eqn:cdobj}
	\end{equation}
	where $\hat{r}_{u, i, ee'} = \hat{r}_{u, i, e} - \hat{r}_{u, i, e'}$ is the score difference between a pair of interactions.
	\item \textbf{PITF}: Pairwise Interaction Tensor Factorization \cite{WSDM10-PITF}.
	It makes prediction for a triplet based on Eq. \eqref{eqn:pitf}, and its objective function is identical to CD's in Eq. \eqref{eqn:cdobj}.
\end{itemize}

\begin{table}
	\centering
	\caption{Example explanations on the three datasets.}
	\begin{tabular}{l}
		\hline \hline
		\multicolumn{1}{c}{\textbf{Amazon Movies \& TV}} \\ \hline
		Great story \\ \hline
		Don't waste your money \\ \hline
		The acting is great \\ \hline
		The sound is okay \\ \hline
		A wonderful movie for all ages \\ \hline
		\hline
		\multicolumn{1}{c}{\textbf{TripAdvisor}} \\ \hline
		Great location \\ \hline
		The room was clean \\ \hline
		The staff were friendly and helpful \\ \hline
		Bad service \\ \hline
		Comfortable hotel with good facilities \\ \hline
		\hline
		\multicolumn{1}{c}{\textbf{Yelp}} \\ \hline
		Great service \\ \hline
		Everything was delicious \\ \hline
		Prices are reasonable \\ \hline
		This place is awful \\ \hline
		The place was clean and the food was good \\ \hline \hline
	\end{tabular}
	\label{tbl:case}
\end{table}

To verify the effectiveness of the joint-ranking framework, in addition to our method BPER-J, we also present the results of two baselines: CD \cite{Springer70-CD} and PITF \cite{WSDM10-PITF}.
Since CD and PITF are not originally designed to accomplish the two tasks of item recommendation and explanation ranking together, we first allow them to make prediction for a user-item pair $(u, i)$ via the inner product of their latent factors, i.e., $\hat{r}_{u, i} = \mathbf{p}_u^T \mathbf{q}_i$, and then combine this task with explanation ranking in a multi-task learning framework whose objective function is given below:
\begin{equation}
	\min_{\Theta} \sum_{u \in \mathcal{U}} \sum_{i \in \mathcal{I}_u} \Big[ \sum_{i' \in \mathcal{I} / \mathcal{I}_u} - \ln \sigma (\hat{r}_{u, ii'}) + \alpha \sum_{e \in \mathcal{E}_{u, i}} \sum_{e' \in \mathcal{E} / \mathcal{E}_{u, i}} - \ln \sigma (\hat{r}_{u, i, ee'}) \Big] + \lambda \left| \left| \Theta \right| \right|_F^2
\end{equation}
where $\hat{r}_{u, ii'} = \hat{r}_{u, i} - \hat{r}_{u, i'}$ is the difference between a pair of records.
We name them CD-J and PITF-J respectively, where J denotes joint-ranking.

\subsection{Evaluation Metrics}

To evaluate the performance of both recommendation and explanation, we adopt four commonly used ranking-oriented metrics in recommender systems: Normalized Discounted Cumulative Gain (\textbf{NDCG}), Precision (\textbf{Pre}), Recall (\textbf{Rec}) and \textbf{F1}.
We evaluate on top-10 ranking for both recommendation and explanation tasks.
For the former task, it is easy to find the definition of the metrics in previous works, so we define those for the latter.
Specifically, the scores for a user-item pair on the four metrics are computed as follows,
\begin{equation}
\begin{aligned}
	\text{rel}_p &= \delta (\text{Top}(u, i, N, p) \in \mathcal{E}_{u, i}^{te}) \\
	\text{NDCG}(u, i, N) & = \frac {1}{Z} \sum_{p = 1}^N \frac {2^{\text{rel}_p} - 1}{\log (p + 1)} \mbox{ where } Z = \sum_{p = 1}^N \frac {1}{\log (p + 1)}  \\
	\text{Pre}(u, i, N) & = \frac {1} {N} \sum_{p = 1}^N \text{rel}_p \mbox{ and } \text{Rec}(u, i, N) = \frac {1} {\left| \mathcal{E}_{u, i}^{te} \right|} \sum_{p = 1}^N \text{rel}_p \\
	\text{F1}(u, i, N) & = 2 \times \frac {\text{Pre}(u, i, N) \times \text{Rec}(u, i, N)} {\text{Pre}(u, i, N) + \text{Rec}(u, i, N)}
\end{aligned}
\end{equation}
where $\text{rel}_p$ indicates whether the $p$-th explanation in the ranked list $\text{Top}(u, i, N)$ can be found in the ground-truth explanation set $\mathcal{E}_{u, i}^{te}$.
Then, we can average the scores for all user-item pairs in the testing set.

\subsection{Implementation Details}

We randomly divide each dataset into training (70\%) and testing (30\%) sets, and guarantee that each user/item/explanation has at least one record in the training set.
The splitting process is repeated for 5 times.
For validation, we randomly draw 10\% records from training set.
After hyper-parameters tuning, the average performance on the 5 testing sets is reported.

We implemented all the methods in Python\footnote{https://www.python.org}.
For TF-based methods, including CD, PITF, CD-J, PITF-J, and our BPER and BPER-J, we search the dimension of latent factors $d$ from [10, 20, 30, 40, 50], regularization coefficient $\lambda$ from [0.001, 0.01, 0.1], learning rate $\gamma$ from [0.001, 0.01, 0.1], and maximum iteration number $T$ from [100, 500, 1000].
As to joint-ranking of CD-J, PITF-J and our BPER-J, the regularization coefficient $\alpha$ on explanation task is searched from [0, 0.1, ..., 0.9, 1].
For the evaluation of joint-ranking, we first evaluate the performance of item recommendation for users, followed by the evaluation of explanation ranking on those correctly predicted user-item pairs.
For our methods BPER and BPER-J, the parameter $\mu$ that balances user and item scores for explanation ranking is searched from [0, 0.1, ..., 0.9, 1].
After parameter tuning, we use $d = 20$, $\lambda = 0.01$, $\gamma = 0.01$ and $T = 500$ for our methods, while the other parameters $\alpha$ and $\mu$ are dependent on the datasets.

The configuration of BPER+ is slightly different, because of the textual content of the explanations.
We adopted the pre-trained BERT from huggingface\footnote{https://huggingface.co/bert-base-uncased}, and implemented the model in Python with PyTorch\footnote{https://pytorch.org}.
We set batch size to 128, $d = 20$ and $T = 5$.
After parameter tuning, we set learning rate $\gamma$ to 0.0001 on Amazon, and 0.00001 on both TripAdvisor and Yelp.

\begin{table*}[!t]
	\centering
	\caption{Performance comparison of all methods on the top-10 explanation ranking in terms of NDCG, Precision, Recall and F1 (\%). The best performing values are boldfaced, and the second best underlined. Improvements are made by BPER+ over the best baseline PITF (* indicates the statistical significance over PITF for $p <$ 0.01 via Student's t-test).}
	\begin{tabular}{l|ccccc}
		\hline
		\multicolumn{1}{c|}{}  & NDCG@10 (\%) & Precision@10 (\%) & Recall@10 (\%) & F1@10 (\%) & Training Time \\ \hline \hline		
		\multicolumn{1}{c|}{\multirow{2}{*}{}} & \multicolumn{5}{c}{\textbf{Amazon Movies \& TV}} \\ \hline
		CD & 0.001 & 0.001 & 0.007 & 0.002 & 1h48min \\
		RAND & 0.004 & 0.004 & 0.027 & 0.006 & - \\
		RUCF & 0.341 & 0.170 & 1.455 & 0.301 & - \\
		RICF & 0.417 & 0.259 & 1.797 & 0.433 & - \\
		PITF & 2.352 & 1.824 & 14.125 & 3.149 & 1h51min \\
		\textbf{BPER} & \underline{2.630}* & \textbf{1.942}* & \textbf{15.147}* & \textbf{3.360}* & 1h56min \\
		\textbf{BPER+} & \textbf{2.877}* & \underline{1.919}* & \underline{14.936}* & \underline{3.317}* & - \\ \hline
		\tiny{Improvement (\%)} & 22.352 & 5.229 & 5.739 & 5.343 & - \\ \hline \hline
		\multicolumn{1}{c|}{\multirow{2}{*}{}} & \multicolumn{5}{c}{\textbf{TripAdvisor}} \\ \hline
		CD & 0.001 & 0.001 & 0.003 & 0.001 & 5h32min \\
		RAND & 0.002 & 0.002 & 0.011 & 0.004 & - \\
		RUCF & 0.260 & 0.151 & 0.779 & 0.242 & - \\
		RICF & 0.031 & 0.020 & 0.087 & 0.030 & - \\
		PITF & 1.239 & 1.111 & 5.851 & 1.788 & 7h9min \\
		\textbf{BPER} & \underline{1.389}* & \underline{1.236}* & \underline{6.549}* & \underline{1.992}* & 9h43min\\
		\textbf{BPER+} & \textbf{2.096}* & \textbf{1.565}* & \textbf{8.151}* & \textbf{2.515}* & - \\ \hline
		\tiny{Improvement (\%)} & 69.073 & 40.862 & 39.314 & 40.665 & - \\ \hline \hline
		\multicolumn{1}{c|}{\multirow{2}{*}{}} & \multicolumn{5}{c}{\textbf{Yelp}} \\ \hline
		CD & 0.000 & 0.000 & 0.003 & 0.001 & 12h7min \\
		RAND & 0.001 & 0.001 & 0.007 & 0.002 & - \\
		RUCF & 0.040 & 0.020 & 0.125 & 0.033 & - \\
		RICF & 0.037 & 0.026 & 0.137 & 0.042 & - \\
		PITF & 0.712 & 0.635 & 4.172 & 1.068 & 11h27min \\
		\textbf{BPER} & \underline{0.814}* & \underline{0.723}* & \textbf{4.768}* & \underline{1.218}* & 16h30min\\
		\textbf{BPER+} & \textbf{0.903}* & \textbf{0.731}* & \underline{4.544}* & \textbf{1.220}* & - \\ \hline
		\tiny{Improvement (\%)} & 26.861 & 15.230 & 8.925 & 14.228 & - \\ \hline \hline
	\end{tabular}
	\label{tbl:exp}
\end{table*}

\section{Results and Analysis} \label{sec:results}

In this section, we first compare our methods BPER and BPER+ with baselines regarding explanation ranking.
Then, we study the capability of our methods in dealing with varying data sparseness.
Third, we show a case study of explanation ranking for both recommendation and disrecommendation,  and also present a small user study.
Lastly, we analyze the joint-ranking results of three TF-based methods.

\begin{figure*}
	\centering
	\subfigure{\includegraphics[scale=0.5]{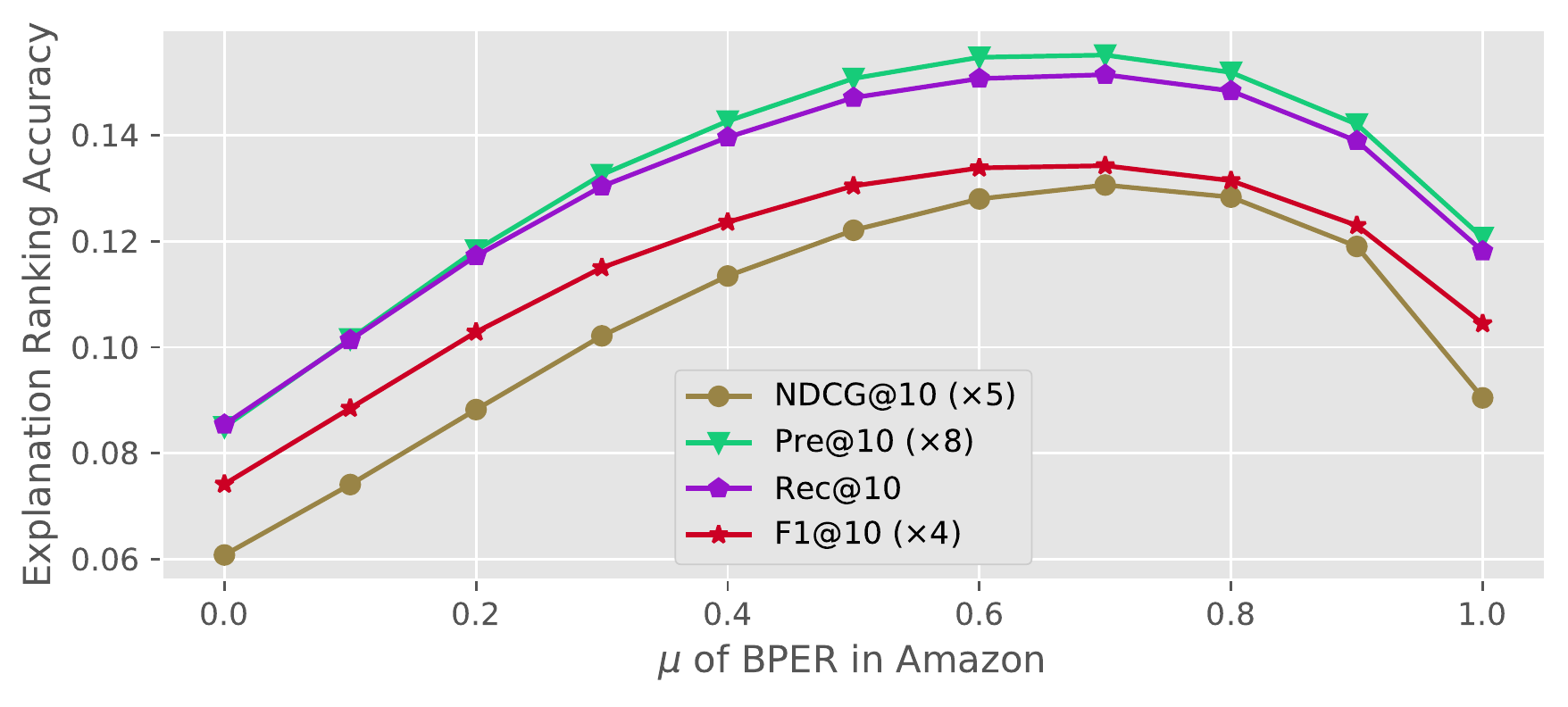}}
	\subfigure{\includegraphics[scale=0.5]{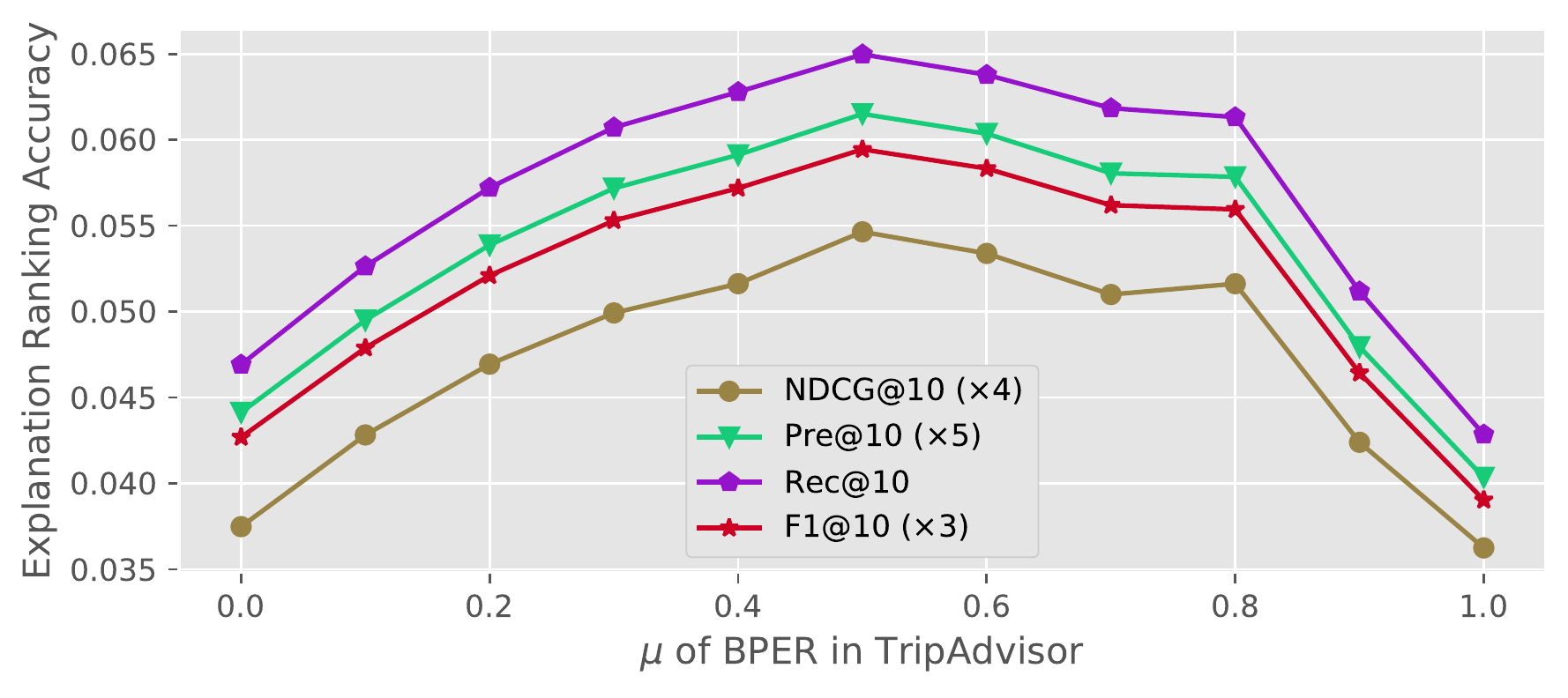}}
	\subfigure{\includegraphics[scale=0.5]{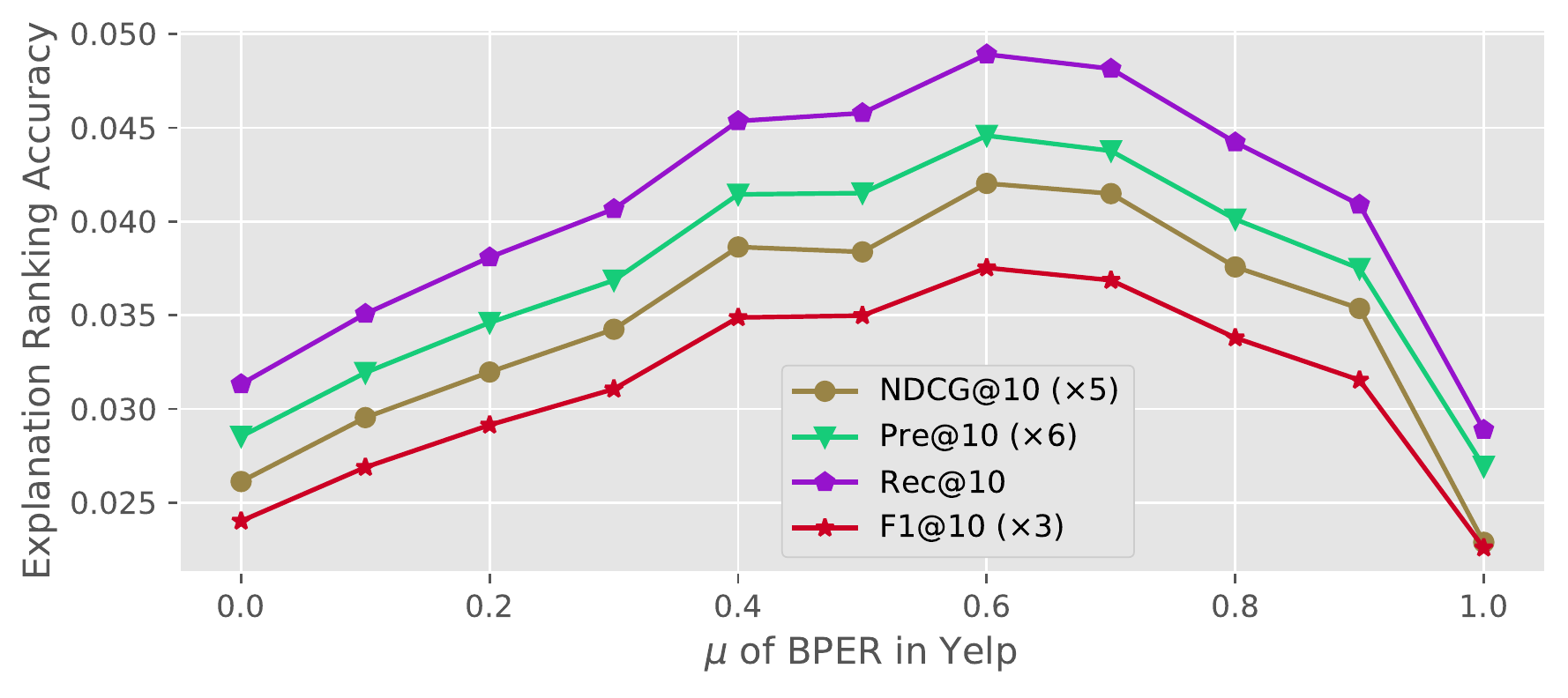}}
	\caption{The effect of $\mu$ in BPER on explanation ranking in three datasets. NDCG@10, Pre@10 and F1@10 are linearly scaled for better visualization.}
	\label{fig:mu}
\end{figure*}

\subsection{Comparison of Explanation Ranking}  \label{sec:mu}

Experimental results for explanation ranking on the three datasets are shown in Table \ref{tbl:exp}.
We see that each method's performance on the four metrics (i.e., NDCG, Precision, Recall, F1) are fairly consistent across the three datasets.
The method RAND is among the weakest baselines, because it randomly selects explanations without considering user and item information, which implies that the explanation ranking task is non-trivial.
CD performs even worse than RAND, because of the sparsity issue in the ternary data (see Table \ref{tbl:dataset}), for which CD may not be able to mitigate as discussed in Section \ref{sec:bper}.
CF-based methods, i.e., RUCF and RICF, largely advance the performance of RAND, as they take into account the information of either users or items, which confirms the important role of personalization for explanation ranking.
However, their performance is still limited due to data sparsity.
PITF and our BPER/BPER+ outperform the CF-based methods by a large margin, as they not only address the data sparsity issue via their MF-like model structure, but also take each user's and item's information into account using latent factors.
Most importantly, our method BPER significantly outperforms the strongest baseline PITF, owing to its ability of producing two sets of scores, corresponding to users and items respectively, and its parameter $\mu$ that can balance their relative importance to explanation ranking.
Lastly, BPER+ further improves BPER on most of the metrics across the three datasets, especially on NDCG that cares about the ranking order, which can be attributed to the consideration of the semantic features of the explanations as well as BERT's strong language modeling capability to extract them.

Besides the explanation ranking performance, we also present the training time comparison of the three TF-based methods in Table \ref{tbl:exp}. For fair comparison, the runtime testing is conducted on the same research machine without GPU, because these methods are all implemented in pure Python without involving deep learning framework. From the table, we can see that the training time of the three methods is generally consistent on different datasets. CD takes the least time to train, PITF needs a bit more training time, while the duration of training our BPER is the longest. This is quite expected since the model complexity grows larger from CD to PITF and BPER. However, the slightly sacrificed training time of BPER is quite acceptable because the gap of training duration between the three methods is not very large, e.g., 5h32min for CD, 7h9min for PITF and 9h43min for BPER on TripAdvisor dataset.

At last, we further analyze the parameter $\mu$ of BPER that controls the contributions of user scores and item scores in Eq. \eqref{eqn:predict}.
As it can be seen in Fig. \ref{fig:mu}, the curves of NDCG, Precision, Recall and F1 are all bell-shaped, where the performance improves significantly with the increase of $\mu$ until it reaches an optimal point, and then it drops sharply.
Due to the characteristics of different application domains, the optimal points vary among the three datasets, i.e., 0.7 for both Amazon and Yelp and 0.5 for TripAdvisor.
We omit the figures of BPER+, because the pattern is similar.

\begin{figure*}
	\centering
	\subfigure{\includegraphics[scale=0.5]{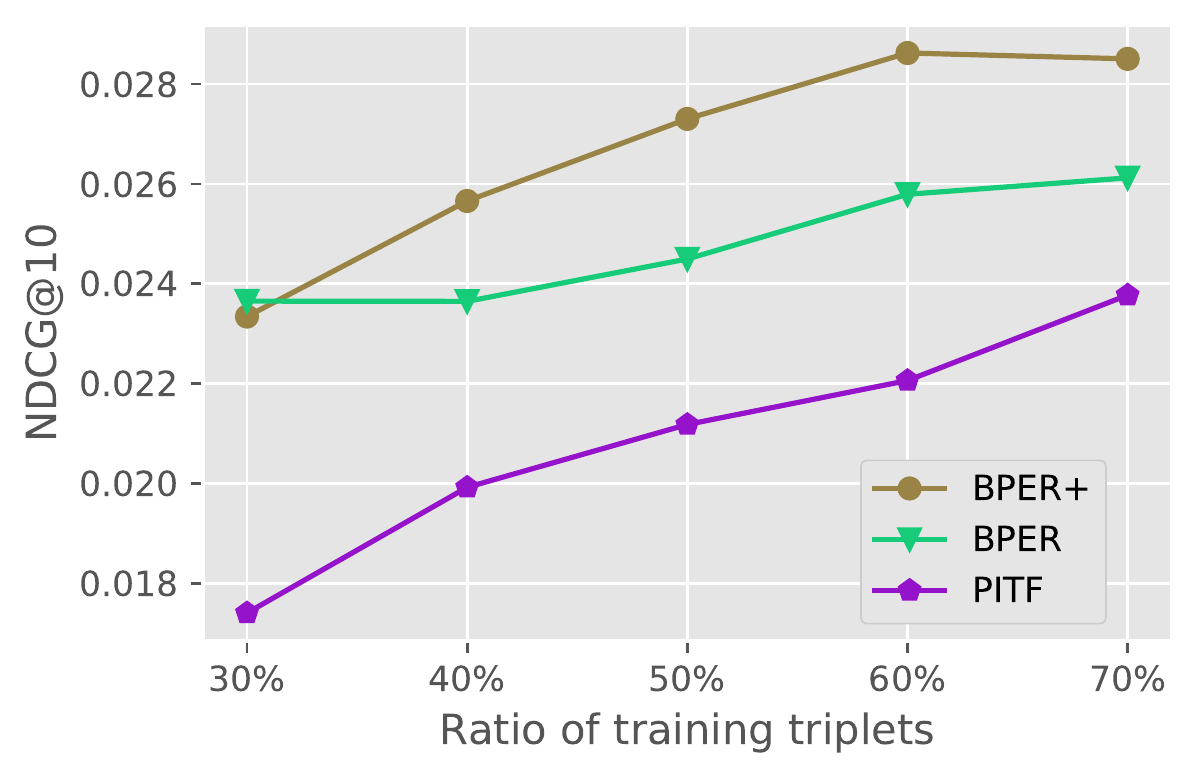}}
	\subfigure{\includegraphics[scale=0.5]{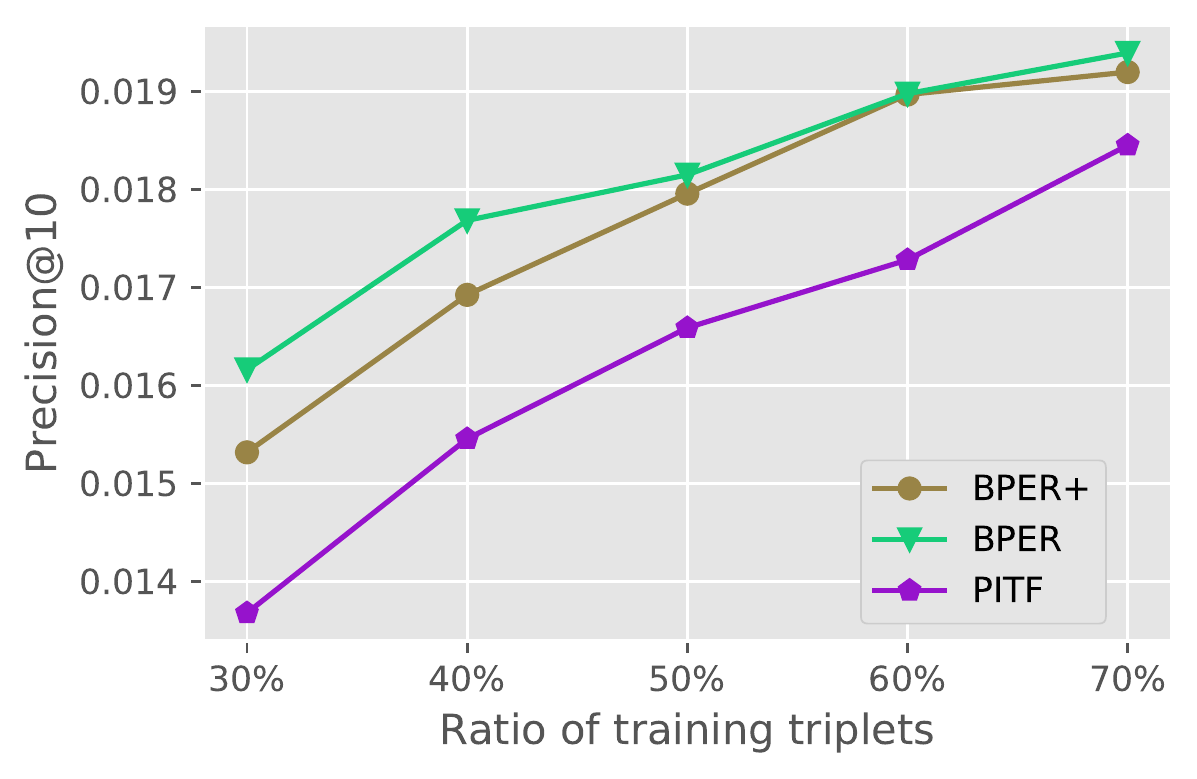}}
	\subfigure{\includegraphics[scale=0.5]{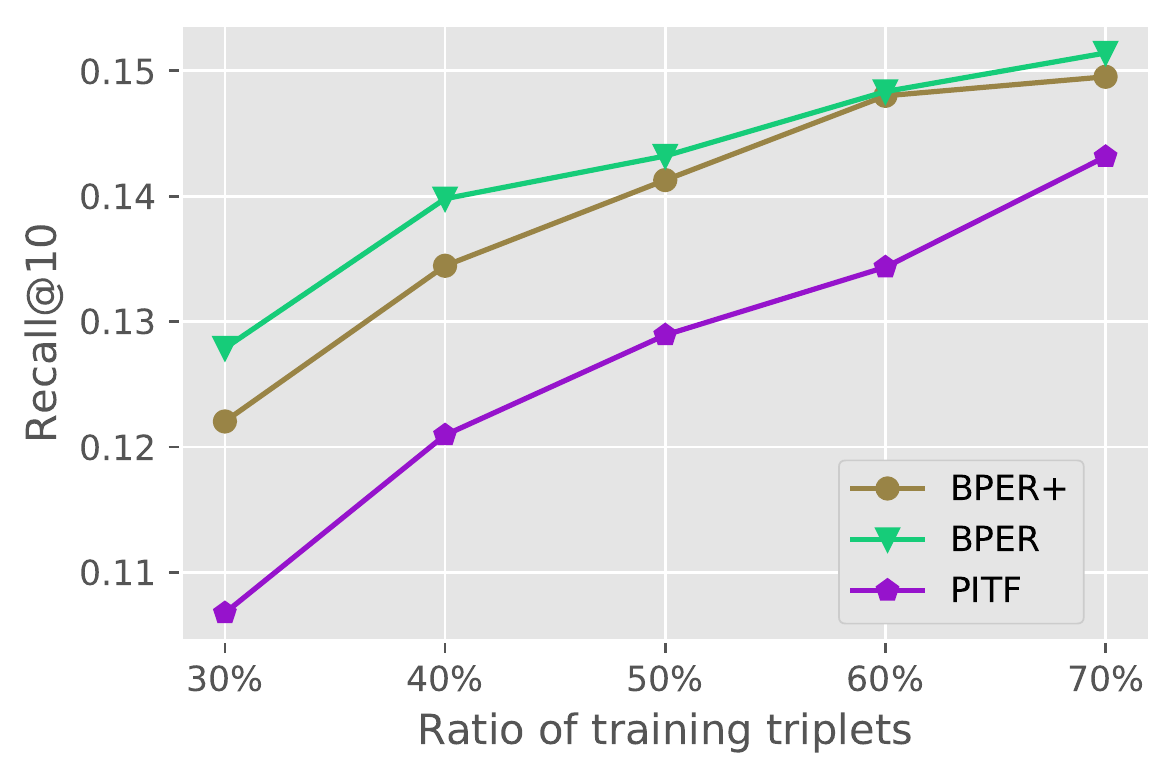}}
	\subfigure{\includegraphics[scale=0.5]{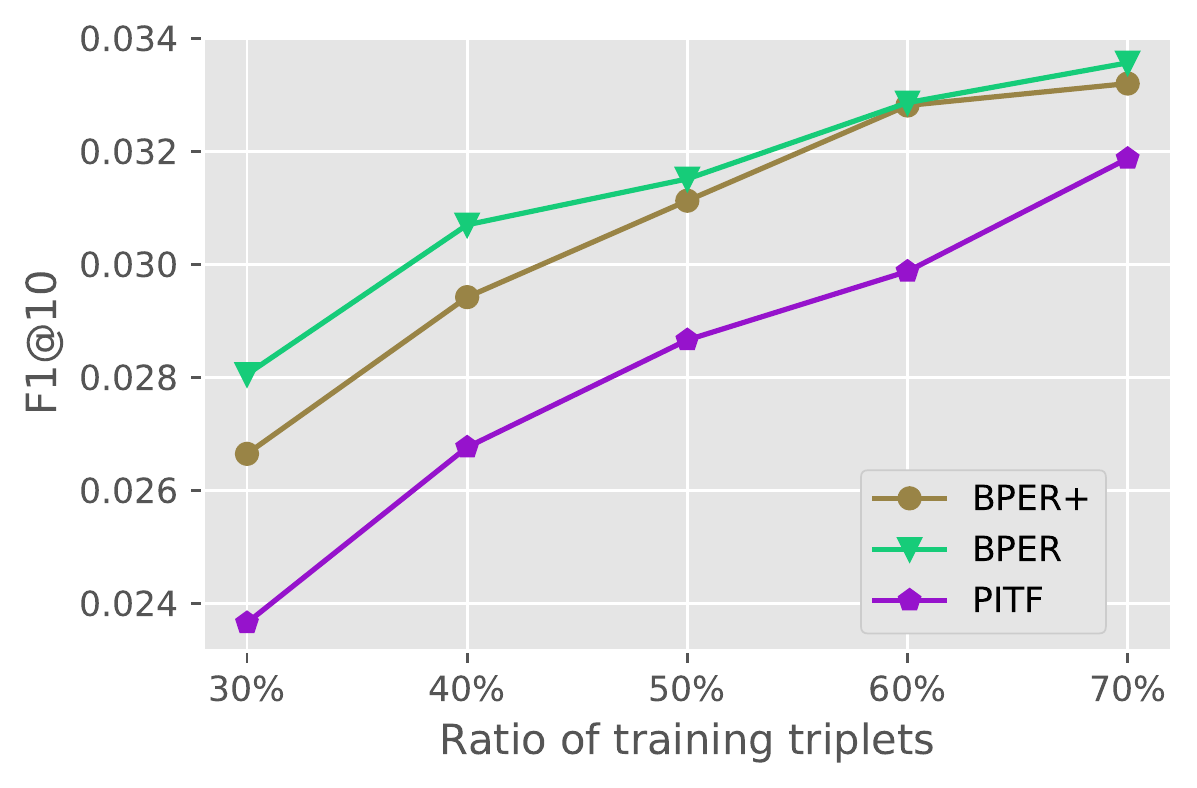}}
	\caption{Ranking performance of three TF-based methods w.r.t. varying sparseness of training data on Amazon dataset.}
	\label{fig:sparsity}
\end{figure*}

\subsection{Results on Varying Data Sparseness}  \label{sec:sparse}

As discussed earlier, the sparsity issue of user-item-explanation triple-wise data is severer than that of traditional user-item pair-wise data.
To investigate how different methods deal with varying spareness, we further remove certain ratio of the Amazon training set, so that the training triplets to the whole dataset ranges from 30\% to 70\%, while the testing set remains untouched.
For comparison with our BPER and BPER+, we include the most competitive baseline PITF.
Fig. \ref{fig:sparsity} shows the ranking performance of the three methods w.r.t. varying spareness.
The ranking results are quite consistent on the four metrics (i.e., NDCG, Precision, Recall and F1).
Moreover, with the increase of the amount of training triplets, the performance of all three methods goes up linearly.
Particularly, the performance gap between our BPER/BPER+ and PITF is quite large, especially when the ratio of training data is small (e.g., 30\%).
These observations demonstrate our methods' better capability in mitigating data sparsity issue, and hence prove the rationale of our solution that converts triplets to two groups of binary relation.

\begin{table}
	\centering
	\caption{Top-5 explanations selected by BPER and PITF for two given user-item pairs, corresponding to recommendation and disrecommendation, on Amazon Movies \& TV dataset. The ground-truth explanations are unordered. Matched explanations are emphasized in italic font.}
	\begin{tabular}{l|l|l}
		\hline \hline
		\textbf{Ground-truth} & \textbf{BPER} & \textbf{PITF} \\ \hline \hline
		Special effects & \textit{Special effects} & Great special effects \\
		Great story & Good acting & Great visuals \\
		Wonderful movie & This is a great movie & Great effects \\
		& \textit{Great story} & \textit{Special effects} \\
		& Great special effects & Good movie \\ \hline
		The acting is terrible & \textit{The acting is terrible} & Good action movie \\
		& The acting is bad & Low budget \\
		& The acting was horrible & Nothing special \\
		& It's not funny & The acting is poor \\
		& Bad dialogue & The acting is bad \\ \hline
	\end{tabular}
	\label{tbl:rankcase}
\end{table}

\begin{figure}
	\centering
	\includegraphics[scale=0.6]{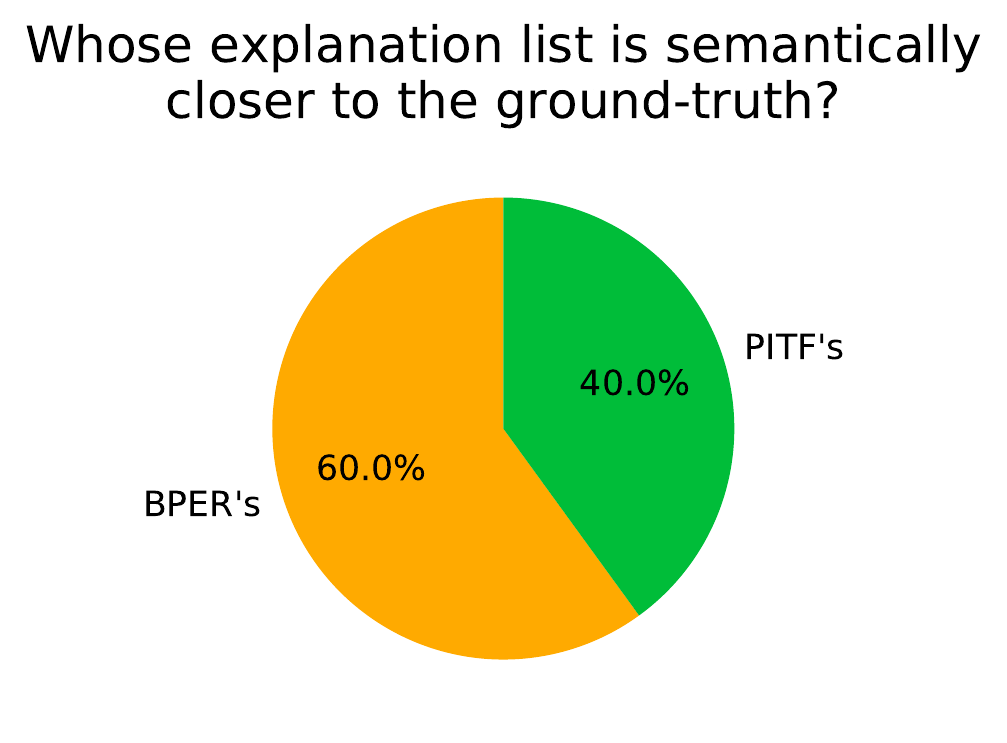}
	\caption{Result of user study on explanations returned by two methods on Amazon Movies \& TV dataset.}
	\label{fig:survey}
\end{figure}

\subsection{Qualitative Case Study and User Study}  \label{sec:case}

To better understand how explanation ranking works, we first present a case study comparing our method BPER and the most effective baseline PITF on Amazon Movies \& TV dataset in Table \ref{tbl:rankcase}.
The two cases in the table respectively correspond to recommendation and disrecommendation.
In the first case (i.e., recommendation), there are three ground-truth explanations, praising the movie's ``\textit{special effects}'', ``\textit{story}'' and overall quality.
Generally speaking, the top-5 explanations resulting from both BPER and PITF are positive, and relevant to the ground-truth, because the two methods are both effective in terms of explanation ranking.
However, since PITF's ranking ability is relatively weaker than our BPER, its explanations miss the key feature ``\textit{story}'' that the user also cares about.

In the second case (i.e., disrecommendation), the ground-truth explanation is a negative comment about the target movie's ``\textit{acting}''.
Although the top explanations made by both BPER and PITF contain negative opinions regarding this aspect, their ranking positions are quite different (i.e., top-3 for our BPER vs. bottom-2 for PITF).
Moreover, we notice that for this disrecommendation, PITF places a positive explanation in the 1st position, i.e., ``\textit{good action movie}'', which not only contradicts the other two explanations, i.e., ``\textit{the acting is poor/bad}'', but also mismatches the disrecommendation goal.
Again, this showcases our model's effectiveness for explanation ranking.

We further conduct a small scale user study to investigate real people's perception towards the top ranked explanations.
Specifically, we still compare our BPER with PITF on Amazon Movies \& TV dataset.
We prepared ten different cases and hired college students to do the evaluation.
In each case, we provide the movie's title and the ground-truth explanations, and ask the participants to select one explanation list that is semantically closer to the ground-truth.
There are two randomly shuffled options returned by BPER and PITF, respectively.
A case is valid only when at least two participants select the same option.
The evaluation results are shown in Fig. \ref{fig:survey}.
We can see that on 60\% of cases our BPER's explanations are closer to the ground-truth than PITF's, which is quite consistent with their explanation ranking performance.

\begin{figure*}
	\centering
	\subfigure{\includegraphics[scale=0.5]{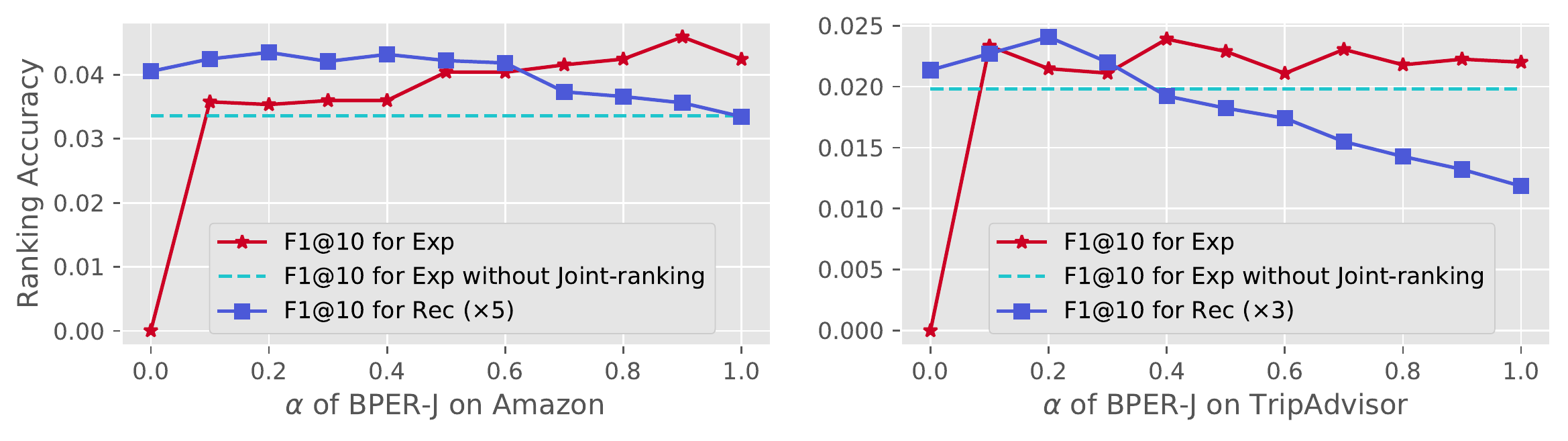}}
	\subfigure{\includegraphics[scale=0.5]{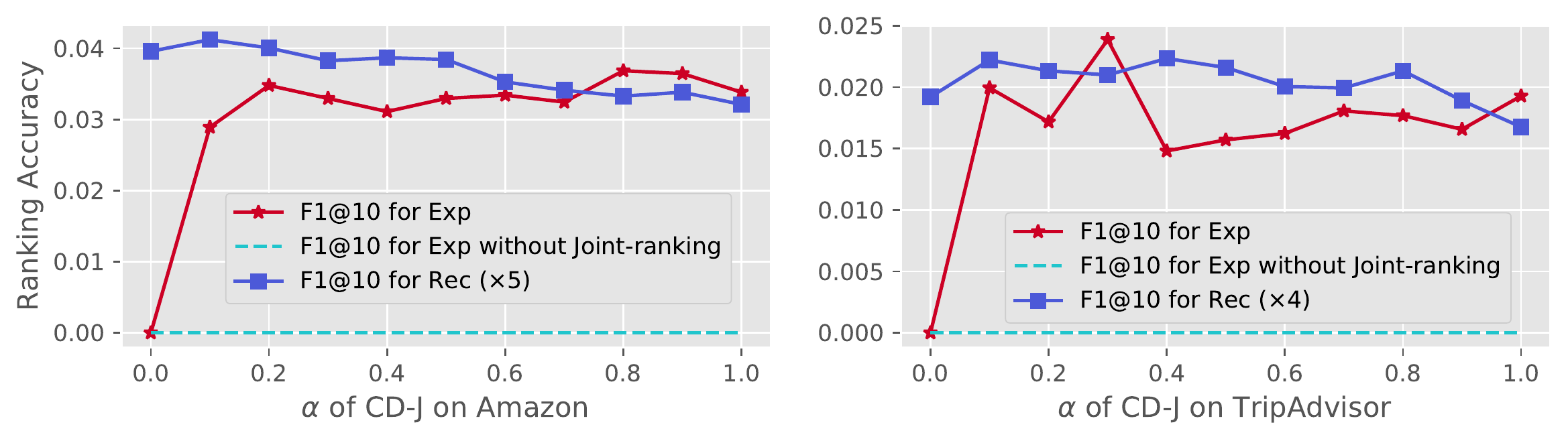}}
	\subfigure{\includegraphics[scale=0.5]{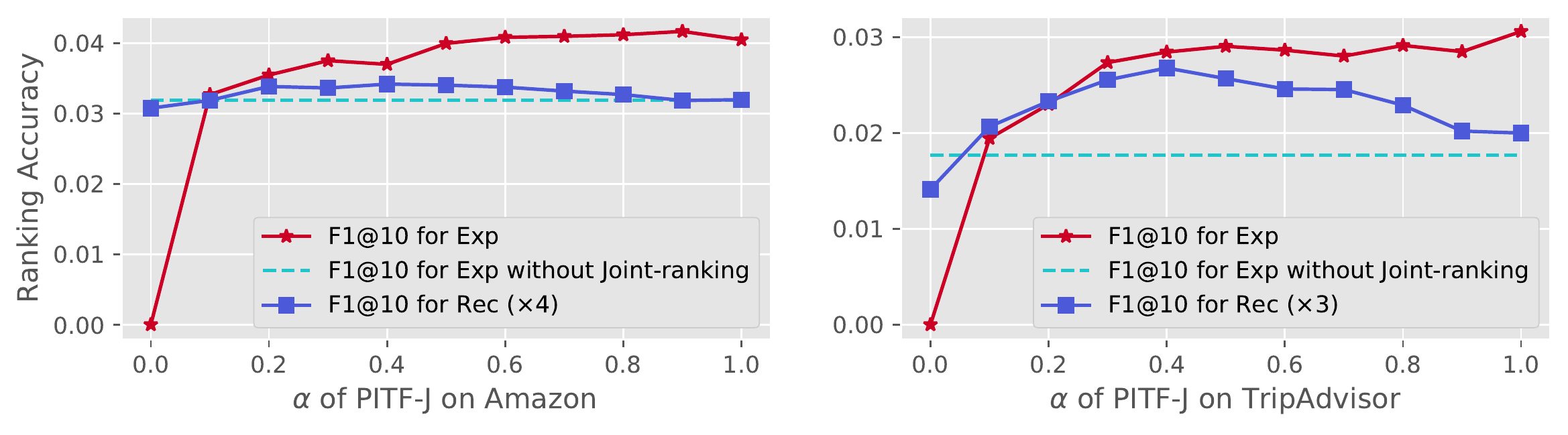}}
	\caption{The effect of $\alpha$ in three TF-based methods with joint-ranking on two datasets. Exp and Rec respectively denote the Explanation and Recommendation tasks. F1@10 for Rec is linearly scaled for better visualization.}
	\label{fig:alpha}
\end{figure*}

\begin{table*}[!t]
	\centering
	\caption{Self-comparison of three TF-based methods on two datasets with and without joint-ranking in terms of NDCG and F1. Top-10 results are evaluated for both explanation (Exp) and recommendation (Rec) tasks. The improvements are made by the best performance of each task under joint-ranking over that without it (i.e., in this case the two tasks are separately learned).}
	\begin{tabular}{l|l|cc|cc|cc|cc}
		\hline
		\multicolumn{2}{c|}{} & \multicolumn{4}{c|}{Amazon} & \multicolumn{4}{c}{TripAdvisor} \\ \cline{3-10}
		\multicolumn{2}{c|}{} & \multicolumn{2}{c|}{Exp (\%)} & \multicolumn{2}{c|}{Rec (\textperthousand)} & \multicolumn{2}{c|}{Exp (\%)} & \multicolumn{2}{c}{Rec (\textperthousand)} \\ \cline{3-10}
		\multicolumn{2}{c|}{} & NDCG & F1 & NDCG & F1 & NDCG & F1 & NDCG & F1 \\ \hline \hline
		\multicolumn{2}{c|}{} & \multicolumn{8}{c}{\textbf{BPER-J}} \\ \hline
		\multicolumn{2}{c|}{Non-joint-ranking} & 2.6 & 3.4 & 6.6 & 8.1 & 1.4 & 2.0 & 5.3 & 7.1 \\ \cline{1-2}
		\multirow{2}{*}{Joint-ranking} & Best Exp & 3.3 $\uparrow$ & 4.6 $\uparrow$ & 5.7 $\downarrow$ & 7.1 $\downarrow$ & 1.6 $\uparrow$ & 2.4 $\uparrow$ & 5.0 $\downarrow$ & 6.4 $\downarrow$ \\
		&Best Rec & 2.6 $\updownarrow$ & 3.5 $\uparrow$ & 7.1 $\uparrow$ & 8.7 $\uparrow$ & 1.5 $\uparrow$ & 2.1 $\uparrow$ & 6.3 $\uparrow$ & 8.0 $\uparrow$ \\
		\hline
		\multicolumn{2}{c|}{Improvement (\%)} & 26.9 & 35.3 & 7.6 & 7.4 & 14.3 & 20.0 & 18.9 & 11.3 \\ \hline \hline
		
		\multicolumn{2}{c|}{} & \multicolumn{8}{c}{\textbf{CD-J}} \\ \hline
		\multicolumn{2}{c|}{Non-joint-ranking} & 0.0 & 0.0 & 6.5 & 7.9 & 0.0 & 0.0 & 4.5 & 4.8 \\ \cline{1-2}
		\multirow{2}{*}{Joint-ranking} & Best Exp & 2.6 $\uparrow$ & 3.7 $\uparrow$ & 5.5 $\downarrow$ & 6.7 $\downarrow$ & 1.7 $\uparrow$ & 2.4 $\uparrow$ & 4.6 $\uparrow$ & 5.2 $\uparrow$ \\
		&Best Rec & 1.9 $\uparrow$ & 2.9 $\uparrow$ & 6.8 $\uparrow$ & 8.2 $\uparrow$ & 9.6 $\uparrow$ & 1.5 $\uparrow$ & 4.9 $\uparrow$ & 5.6 $\uparrow$ \\
		\hline
		\multicolumn{2}{c|}{Improvement (\%)} & Inf & Inf & 4.6 & 3.8 & Inf & Inf & 8.9 & 16.7 \\ \hline \hline
		
		\multicolumn{2}{c|}{} & \multicolumn{8}{c}{\textbf{PITF-J}} \\ \hline
		\multicolumn{2}{c|}{Non-joint-ranking} & 2.4 & 3.2 & 6.5 & 7.7 & 1.2 & 1.8 & 4.3 & 4.7 \\ \cline{1-2}
		\multirow{2}{*}{Joint-ranking} & Best Exp & 3.0 $\uparrow$ & 4.2 $\uparrow$ & 6.4 $\downarrow$ & 8.0 $\uparrow$ & 2.0 $\uparrow$ & 2.9 $\uparrow$ & 6.0 $\uparrow$ & 7.6 $\uparrow$ \\
		&Best Rec & 2.8 $\uparrow$ & 3.7 $\uparrow$ & 7.1 $\uparrow$ & 8.5 $\uparrow$ & 2.0 $\uparrow$ & 2.8 $\uparrow$ & 7.0 $\uparrow$ & 8.9 $\uparrow$ \\
		\hline
		\multicolumn{2}{c|}{Improvement (\%)} & 25.0 & 31.3 & 9.2 & 10.4 & 66.7 & 61.1 & 62.8 & 89.4 \\ \hline \hline
		
	\end{tabular}
	\label{tbl:corank}
\end{table*}

\subsection{Effect of Joint-Ranking}

We perform joint-ranking for three TF-based models, i.e., BPER-J, CD-J and PITF-J.
Because of the consistency in the experimental results on different datasets, we only show results on Amazon and TripAdvisor.
In Fig. \ref{fig:alpha}, we study the effect of the parameter $\alpha$ to both explanation ranking and item ranking in terms of F1 (results on the other three metrics are consistent).
In each sub-figure, the green dotted line represents the performance of explanation ranking task without joint-ranking, whose value is taken from Table \ref{tbl:exp}.
As we can see, all the points on the explanation curve (in red) are above this line when $\alpha$ is greater than 0, suggesting that the explanation task benefits from the recommendation task under the joint-ranking framework.
In particular, the explanation performance of CD-J improves dramatically under the joint-ranking framework, since its recommendation task suffers less from the data sparsity issue than the explanation task as discussed in Section \ref{sec:bper}.
It in turn helps to better rank the explanations.
Meanwhile, for the recommendation task, all the three models degenerate to BPR when $\alpha$ is set to 0.
Therefore, on the recommendation curves (in blue), any points, whose values are greater than that of the starting point, gain profits from the explanation task as well.
All these observations show the effectiveness of our joint-ranking framework in terms of enabling the two tasks to benefit from each other.

In Table \ref{tbl:corank}, we make a self-comparison of the three methods in terms of NDCG and F1 (the other two metrics are similar).
In this table, ``Non-joint-ranking'' corresponds to each model's performance with regard to explanation or recommendation when the two tasks are individually learned.
In other words, the explanation performance is taken from Table \ref{tbl:exp}, and the recommendation performance is evaluated when $\alpha = 0$.
``Best Exp'' and ``Best Rec'' denote the best performance of each method on respectively explanation task and recommendation task under the joint-ranking framework.
As we can see, when the recommendation performance is the best for all the models with joint-ranking, the explanation performance is always improved.
Although minor recommendation accuracy is sacrificed when the explanation task reaches the best performance, we can always find points where both of the two tasks are improved, e.g., on the top left of Fig. \ref{fig:alpha} when $\alpha$ is in the range of 0.1 to 0.6 for BPER-J on Amazon.
This again demonstrates our joint-ranking framework's capability in finding good solutions for both tasks.

\section{Conclusion and Future Work} \label{sec:conclude}

To the best of our knowledge, we are the first one that leverages standard offline metrics to evaluate explainability for explainable recommendation.
We achieve this goal by formulating the explanation problem as a ranking task.
With this quantitative measure of explainability, we design an item-explanation joint-ranking framework that can improve the performance of both recommendation and explanation tasks.
To enable such joint-ranking, we develop two effective models to address the data sparsity issue, which were tested on three large datasets.

As future work, we are interested in considering the relationship (such as coherency \cite{IJCAI20-SEER} and diversity) between suggested explanations to further improve the explainability.
In addition, we plan to conduct experiments in real-world systems to validate whether recommendations and their associated explanations as produced by the joint-ranking framework could influence users' behavior, e.g., clicking and purchasing.
Besides, the joint-ranking framework in this paper aims to improve the recommendation performance by providing explanations, while in the future, we will also consider improving other objectives based on explanations, such as recommendation serendipity \cite{WWW19-Serendipity} and fairness \cite{singh2018fairness}.

\begin{acks}
This work was supported by Hong Kong RGC/GRF (RGC/HKBU12201620) and partially supported by NSF IIS-1910154, 2007907, and 2046457. 
Any opinions, findings, conclusions or recommendations expressed in this material are those of the authors and do not necessarily reflect those of the sponsors.
\end{acks}

\bibliographystyle{ACM-Reference-Format}
\bibliography{bibliography}

\appendix

\end{document}